\documentclass{article}

\PassOptionsToPackage{round}{natbib}
\usepackage[preprint]{neurips_2023}

\usepackage[utf8]{inputenc} 
\usepackage[T1]{fontenc}    
\usepackage{hyperref}       
\usepackage{url}            
\usepackage{booktabs}       
\usepackage{amsfonts}       
\usepackage{nicefrac}       
\usepackage{microtype}      
\usepackage{xcolor}         

\usepackage{makecell}
\usepackage{adjustbox}
\usepackage{mathtools}
\usepackage{amsmath}
\usepackage{caption}
\usepackage{subcaption}
\usepackage{lipsum}
\usepackage{wrapfig}
\usepackage{graphicx}

\usepackage{dsfont}

\usepackage{amssymb}
\usepackage{pifont}
\DeclareMathOperator*{\argmin}{arg\,min}

\title{DDDM-VC: Decoupled Denoising Diffusion Models with Disentangled Representation and Prior Mixup for Verified Robust Voice Conversion}
\author{%
    Ha-Yeong Choi$^1$\thanks{Equal contribution}
    \And
    Sang-Hoon Lee$^{1*}$
    \And
    Seong-Whan Lee$^1$\thanks{Corresponding author}
    \AND
    \\
    $^1$Department of Artificial Intelligence, Korea University, Seoul, Korea\\
    \texttt{\{hayeong, sh\_lee, sw.lee\}@korea.ac.kr}\\
}
\begin{document} 
\maketitle
 
\begin{abstract}
Diffusion-based generative models have exhibited powerful generative performance in recent years. However, as many attributes exist in the data distribution and owing to several limitations of sharing the model parameters across all levels of the generation process, it remains challenging to control specific styles for each attribute. To address the above problem, this paper presents decoupled denoising diffusion models (DDDMs) with disentangled representations, which can control the style for each attribute in generative models. We apply DDDMs to voice conversion (VC) tasks to address the challenges of disentangling and controlling each speech attribute (e.g., linguistic information, intonation, and timbre). First, we use a self-supervised representation to disentangle the speech representation. Subsequently, the DDDMs are applied to resynthesize the speech from the disentangled representations for denoising with respect to each attribute. Moreover, we also propose the prior mixup for robust voice style transfer, which uses the converted representation of the mixed style as a prior distribution for the diffusion models. The experimental results reveal that our method outperforms publicly available VC models. Furthermore, we show that our method provides robust generative performance regardless of the model size. Audio samples are available \footnote{\url{https://hayeong0.github.io/DDDM-VC-demo/}}. 
\end{abstract}
 
\section{Introduction} 
Denoising diffusion models \citep{ho2020denoising,dhariwal2021diffusion,song2021scorebased} have achieved significant success in image generation tasks \citep{ramesh2022hierarchical,saharia2022photorealistic}. Diffusion models have also attracted increasing interest in the audio domain in recent years, owing to their ability to synthesize high-quality speech (e.g., Mel-spectrogram and audio). Various applications employ diffusion models, such as text-to-speech (TTS) \citep{popov2021grad,kim2022guided,kim2022guided2}, neural vocoder \citep{kong2021diffwave,chen2021wavegrad,huang2022fastdiff}, speech enhancement \citep{han2022nu}, and voice conversion (VC) \citep{liu2021diffsvc,popov2022diffusionbased}. 

Although diffusion models have achieved success in most speech applications owing to their powerful generative performance, there remains room for improvement in conventional diffusion models. As data include many attributes, it is difficult to control specific styles for each attribute with a single denoiser that shares the model parameters across all levels of generation process. To reduce this burden in the image generation domain, eDiff-i \citep{balaji2022ediffi} subdivides the single denoiser into multiple specialized denoisers that originate from the single denoiser progressively according to specific iterative steps. However, a limitation still exists in controlling each attribute within entirely the same conditioning framework for every iteration, which results in a lack of controllability. 

\begin{wrapfigure}{r}{0.5\columnwidth}
    \centering\vspace{-0.4cm}
     \resizebox{0.5\columnwidth}{!}{
    {\includegraphics[width=0.5\columnwidth]{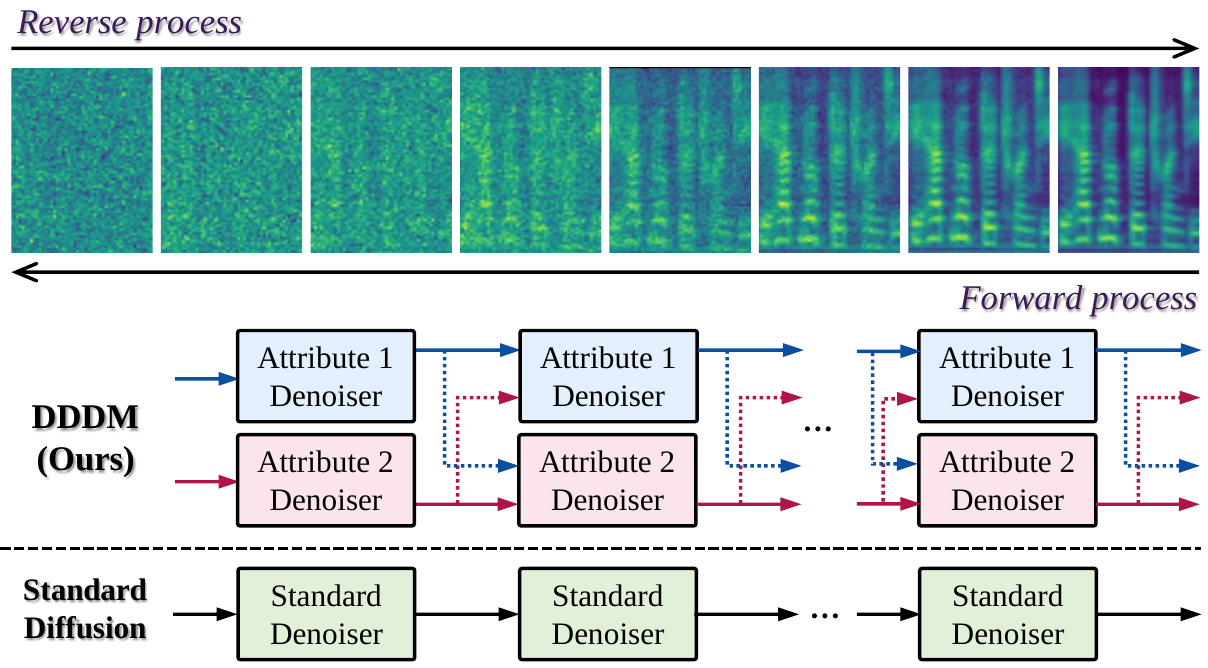}}
    }
    \caption{Speech synthesis in DDDMs and the standard diffusion model. Whereas a single denoiser is used for all denoising steps in standard diffusion models, we subdivide the denoiser into multiple denoisers for each attribute using a self-supervised representation. Each denoiser focuses on removing the single noise from its own attribute in each intermediate time step.}
    \label{fg1}\vspace{-0.3cm}
\end{wrapfigure}

To address the above issues, we first present decoupled denoising diffusion models (DDDMs) with disentangled representations. As illustrated in Figure \ref{fg1}, we disentangle the denoiser into specific attribute-conditioned denoisers to improve the model controllability for each attribute. Subsequently, each denoiser focuses on the noise from its own attribute at the same noise level and removes the noise at each intermediate time step. To demonstrate the effectiveness of DDDMs, we focus on the VC tasks that still face challenges in disentangling and controlling each speech attribute \citep{choi2021neural}. VC is a task for transferring or controlling the voice style while maintaining the linguistic information. As speech consists of various attributes such as linguistic information, intonation, rhythm, and timbre, it remains challenging to transfer the voice style in zero/few-shot scenarios \citep{lee2022hierspeech}. 

Based on the DDDMs, we present DDDM-VC which can effectively transfer and control the voice style for each attribute. We first utilize the self-supervised representation to disentangle the speech representation based on the source-filter theory \citep{fant1970acoustic}. Subsequently, we resynthesize the speech for each attribute from the disentangled representation using DDDMs. We also propose the prior mixup, a novel verified robust voice style transfer training scenario that uses the converted speech as a prior distribution for the diffusion model that is generated from the mixed speech representation, and restores the source speech. Thus, although DDDM-VC is trained by reconstructing the source speech, the prior mixup can reduce the train-inference mismatch problem for VC tasks. We demonstrate that DDDMs can effectively transfer the voice style even with lower model parameters compared to the state-of-the-art VC model \citep{popov2022diffusionbased}. Furthermore, the experimental results reveal the effectiveness of speaker adaptation in the zero/one-shot scenarios. The main contributions of this study are as follows:   
\begin{itemize}
\item We propose decoupled denoising diffusion models (DDDMs), which can effectively control the style for each attribute in generative models by decoupling attributes and adopting the disentangled denosiers. 
\item To demonstrate the effectiveness of DDDMs, We present DDDM-VC, which can disentangle and resynthesize speech for each attribute with self-supervised speech representation. Furthermore, we propose a prior mixup to improve voice style transfer performance. 
\item Our model provides better performance in both many-to-many and zero-shot voice style transfer compared with the state-of-the-art VC model. We can also successfully adapt to novel voice with a single sample. 
\end{itemize}


\section{Related Works: Voice Conversion}

The aim of VC is to convert the source speaker voice into the target speaker voice while preserving the linguistic (content) information \citep{Yi2020}. For this purpose, it is critical to decompose the linguistic, timbre, and pitch information such as intonation. Many VC methods have been presented with the goal of disentangling the speech representation by decomposing the speech into various components: two methods: (1) information bottleneck and (2) information perturbation. 

An information bottleneck is used to constrict the information flow through a carefully designed bottleneck. AutoVC \citep{qian2019autovc} presented a method for inducing the disentanglement of the content and timbre by restricting the dimension of the latent vector between the encoder and decoder. Subsequently, F0-AutoVC \citep{qian2020f0} takes the idea from the constraints of layer dimension \citep{qian2019autovc}, conditioning the decoder with normalized F0. \citep{pmlr-v139-qian21b,lee2021voicemixer} presents a similarity-based information bottleneck for content and style disentanglement. However, in these models, the heuristic determination of appropriate bottleneck size is inevitable, which is directly related to the model performance, and it may even differ for each dataset.       

Information perturbation approaches have been proposed to overcome the above limitation. The basic concept of information perturbation is to remove unnecessary information for each speech representation through signal processing before feeding it to the network. SpeechFlow \citep{qian2020unsupervised} and NANSY \citep{choi2021neural} adopt perturbation methods for the input waveforms and encourage the encoded feature to be correctly removed so that only content information remains. 

Recently, \citep{lee2022duration,popov2022diffusionbased} extracted speaker-irrelevant linguistic information using phoneme information. These models can simply execute content information disentanglement, and convert the speech with accurate pronunciation using explicit phoneme information; however, phoneme information must be extracted in advance and phoneme-level downsampling may induce the loss of content information.

Recent studies have utilized a self-supervised speech representation as the linguistic content representation, \citep{polyak21_interspeech,choi2021neural,huang2021any,huang2022s3prl} for the disentanglement of speech components. Despite the significant advances in VC, there are still certain limitations. The information loss that occurs when disentangling speech representations results in degradation of the synthesized speech quality in terms of both the audio and speaker adaptation quality. Therefore, in this study, we focus on synthesizing high-quality speech by disentangling speech representations appropriately and restoring lost information by using diffusion models. Although \citep{popov2022diffusionbased} employed diffusion models to VC tasks, their method still exhibited limitations in pronunciation and speaker adaptation quality of converted speech. To solve these problems, we present DDDMs and the prior mixup. The details of our methods are described in the following sections.   
\section{Decoupled Denoising Diffusion Models}
\subsection{Background: Diffusion Models}
\label{3-1}

Denoising diffusion models have significantly improved various generative tasks such as image generation \citep{ramesh2022hierarchical,rombach2022high}, image inpainting \citep{saharia2022palette,lugmayr2022repaint}, and audio generation \citep{chen2021wavegrad,kong2021diffwave,huang2022fastdiff}. These models typically consist of a forward process that gradually adds random noise, and a reverse process that progressively removes random noise and restores the original sample. 

 Unlike the original diffusion model that uses a discrete-time diffusion process by Markov chains \citep{ho2020denoising}, the score-based generative model uses a stochastic differential equation (SDE)-based continuous-time diffusion process \citep{song2021scorebased}.
The stochastic forward process is defined as follows:
\begin{equation}
\label{sde1}
d\mathbf{x}=f(\mathbf{x}, t) dt+g(t) d\mathbf{w},
\end{equation}
where $f(., t)$ is the drift coefficient of $\mathbf{x}(t)$, $g(\mathrm{t})$ is the diffusion coefficient, and $\mathbf{w}$ denotes the Brownian motion.
The reverse-time SDE can be expressed as:
\begin{equation}
\label{sde2}
    d\mathbf{x}=[f(\mathbf{x}, t)-g^2(t){\nabla_\mathbf{x} \log p_t(\mathbf{x})}] dt + g(t)d \bar{\mathbf{w}},
\end{equation}
where $\bar{\mathbf{w}}$ is the Brownian motion for the time flowing in backward, and $\nabla_\mathbf{x} \log p_t(\mathbf{x})$ represents the score-function. To estimate $\mathbf{s}_\theta(\mathbf{x}, t)\simeq\nabla_{\mathbf{x}} \log p_t(\mathbf{x})$, the score-based diffusion model is trained with score matching objective:
\begin{equation}
\label{sde3}
   \theta^* = \argmin_\theta \mathds{E}_{t}\Big\{\lambda_{t} \mathds{E}_\mathbf{x_{0}}\mathds{E}_{\mathbf{x_{t}}\mid\mathbf{x_{0}}} \big[\Vert{s_{\theta}(\mathbf{x}_{t}, t) - \nabla_{\mathbf{x}_{t}}\log p_{t|0}(\mathbf{x}_{t}\mid\mathbf{x_{0}})}\Vert_{2}^{2} \big]\Big\}. 
\end{equation}
\vspace{-0.5cm}
\subsection{Disentangled Denoiser}
\label{3-2}
To effectively control the style for each attribute in generative models, we propose decoupled denoising diffusion models (DDDMs) with multiple disentangled denoisers. Although an ensemble of diffusion models was presented in \citep{balaji2022ediffi}, only a single expert is used at the specific denoising step in this method. In contrast, we investigate the decomposition of diffusion models in a single denoising step. Specifically, more than one attribute denoiser is used at any given point.
\begin{figure*}[t]
    \centering
    {\includegraphics[width=1\textwidth]{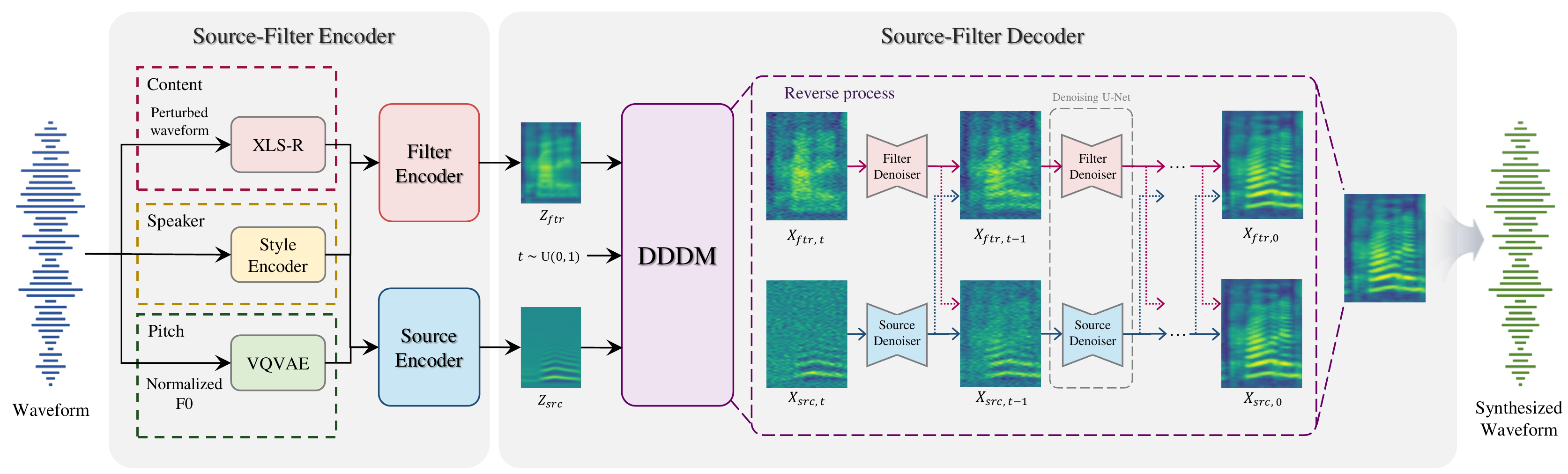}}
    \caption{Overall framework of DDDM-VC}
    \label{fg2}
\end{figure*}
Unlike the general diffusion process, which employs a single denoiser, we subdivide the denoiser into $N$ denoisers with disentangled representations. Following the use of data-driven priors in \citep{popov2022diffusionbased}, we use a disentangled representation of an attribute $Z_n$ as the prior for each attribute denoiser. Therefore, the forward process can be expressed:
\begin{equation}
\label{dd_fwd}
    dX_{n, t}=\frac{1}{2}\beta_{t}({Z}_{n} - X_{n,t})dt + \sqrt{\beta_{t}}d{W_{t}}\ ,
\end{equation}
where $n\in[1,N]$, ${n}$ denotes each attribute, ${N}$ is the total number of attributes, $\beta_{t}$ regulates the amount of stochastic noise and ${W_{t}}$ is the forward Brownian motion.
Reverse trajectories exist for the given forward SDE of each attribute (\ref{dd_fwd}). The reverse process of each disentangled denoiser can be defined as follows:
\begin{equation}
\label{dd_rev}
\begin{split}
    d\hat{X}_{n, t} = {\small\left(\frac{1}{2}({Z_{n}}-\hat{X}_{n,t})-  \sum_{n=1}^{N} s_{\theta_{n}}(\hat{X}_{n,t},{Z_{n}}, t) \right)}\beta_{t}dt +\sqrt{\beta_{t}}d\bar{W_{t}},
\end{split}
\end{equation}
where $t\in[0,1]$, $s_{\theta_{n}}$ represents the score function of each attribute $n$ parameterized by $\theta_{n}$ and $\bar{W_{t}}$ denotes the backward Brownian motion. 
The forward process (\ref{dd_fwd}) that generates a noisy sample $X_{n,t}$ with each prior attribute ${n}$ is as follows: 
\begin{equation} 
\label{dd_fwd_sol}
\begin{split}
    p_{t|0}(X_{n, t}|X_{0}) =  \mathcal{N}\left(e^{-\frac{1}{2}\int_{0}^{t}{\beta_{s}ds}}X_{0}+\left(1-e^{-\frac{1}{2}\int_{0}^{t}{\beta_{s}ds}}\right)Z_{n},
     \left(1-e^{-\int_{0}^{t}{\beta_{s}ds}}\right){\mathrm{I}}\right),
\end{split}
\end{equation}
where $\mathrm{I}$ is the identity matrix. The distribution (\ref{dd_fwd_sol}) is Gaussian, thus we have the following equation:
\begin{equation}
\label{dd_score}
\begin{split}
    \nabla\log{p_{t|0}(X_{n, t}|X_{0})} =  -\frac{X_{n, t}-X_{0}(e^{-\frac{1}{2}\int_{0}^{t}{\beta_{s}ds}})-{Z_{n}}(1-e^{-\frac{1}{2}\int_{0}^{t}{\beta_{s}ds}})}{1 - e^{-\int_{0}^{t}{\beta_{s}ds}}}. 
\end{split}
\end{equation}
The reverse process (\ref{dd_rev}) is trained by optimizing the parameter ${\theta_{n}}$ using the following objective:
\begin{equation}
\label{dd_rev_train} 
\begin{split}
\theta^*_{n} = \argmin_{\theta_{n}}\int_{0}^{1}\lambda_{t}
      \mathds{E}_{X_{0}, X_{n, t}}  \bigg\Vert \sum_{n=1}^{N} s_{\theta_{n}}(X_{n,t}, Z_{n}, s, t) - \nabla{\log{p_{t|0}(X_{n,t}|X_{0})}}\bigg\Vert_{2}^{2}dt, 
\end{split}
\end{equation}
where $\theta=[\theta_1, \cdots, \theta_{N}]$ and $\lambda_{t}=1 - e^{-\int_{0}^{t}{\beta_{s}ds}}$. Furthermore, we derive fast sampling using the ML-SDE solver \citep{popov2022diffusionbased}, which maximizes the log-likelihood of forward diffusion with the reverse SDE solver. We extend DDDMs to DDDM-VC to control the voice style for each attribute in the following Section. In addition, we show that DDDMs can be applied to audio mixing by leveraging multiple denoisers to blend the sound and speech with the desired balance in Appendix H.  

\section{DDDM-VC}
DDDM-VC consists of a source-filter encoder and source-filter decoder as illustrated in Figure \ref{fg2}. We first disentangle the speech using self-supervised speech representations as in subsection \ref{4-1}. Thereafter, we use these disentangled speech representations to control each attribute and to generate high-quality speech with the proposed disentangled denoiser as explained in subsection \ref{4-2}. Furthermore, we propose the prior mixup for a robust voice conversion scenario in subsection \ref{4-3}. 
\subsection{Speech Disentanglement}
\label{4-1}
\paragraph{Content Representation}
To extract the content representation relating to the phonetic information, we utilize self-supervised speech representations. Unlike \citep{polyak21_interspeech} utilizing the discrete representation of audio from HuBERT, we use a continuous representation of audio from XLS-R, which is Wav2Vec 2.0 trained with a large-scale cross-lingual speech dataset. Furthermore, before fed to the filter encoder, audio is perturbed to remove the content-independent information following \citep{choi2021neural}. As \citep{lee2022hierspeech} demonstrated that the representation from the middle layer of XLS-R contains substantial linguistic information, we adopt this representation as the content representation.  

\paragraph{Pitch Representation}
Following \citep{polyak21_interspeech}, we extract the fundamental frequency (F0) from the audio using YAPPT algorithm \citep{kasi2002yet} to encode the intonation such as the speaker-irrelevant speaking style. The F0 from each sample is normalized for each speaker for speaker-independent pitch information, and VQ-VAE is used to extract the vector-quantized pitch representation. For a fair comparison, we normalize the F0 for each sentence, not for a speaker, during inference.

\paragraph{Speaker Representation}
VC transfers the voice style, and our goal is to achieve robust zero-shot voice style transfer from novel speakers. To this end, we use style encoder \citep{min2021meta} that can extract the speaker representation from the Mel-spectrogram of the target speech. The extracted speaker representation is averaged per sentence for global speaker representation, and fed to all encoders and decoders for the speaker adaptation.   

\begin{figure*}
    \centering
    \begin{subfigure}[t]{0.31\textwidth}
    \includegraphics[width=1\textwidth]{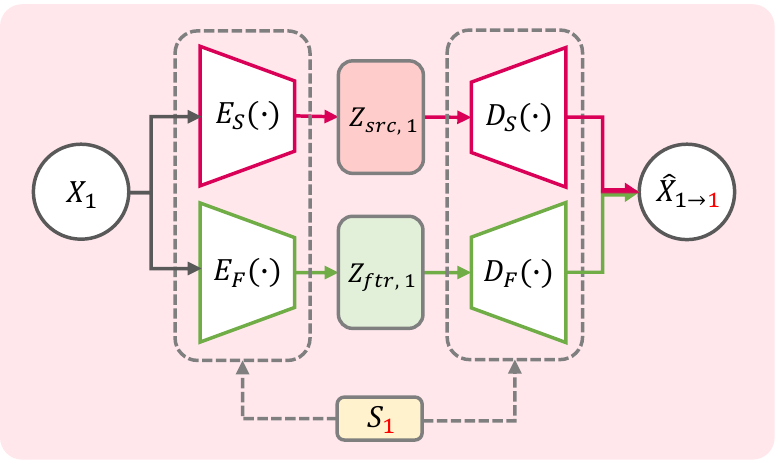}
    \caption{Resynthesis}
    \label{fg3-(a)}
    \end{subfigure}
    \begin{subfigure}[t]{0.31\textwidth}
    \includegraphics[width=1\textwidth]{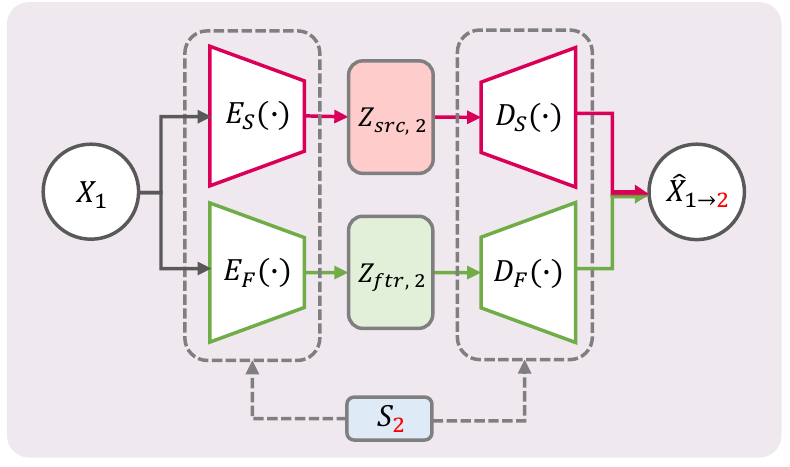}
    \caption{Conversion}
    \label{fg3-(b)}
    \end{subfigure}
    \begin{subfigure}[t]{0.31\textwidth}
    \includegraphics[width=1\textwidth]{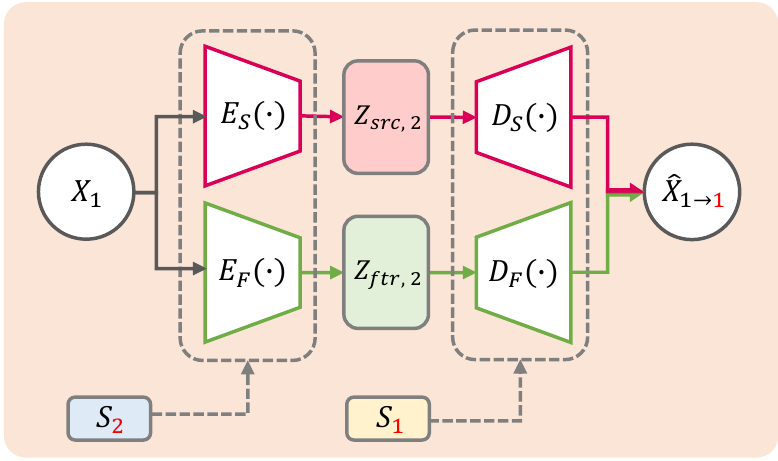}
    \caption{Prior Mixup}
    \label{fg3-(c)}
    \end{subfigure}
\caption{(a) Speech resynthesis from disentangled speech representations (training). (b) Voice conversion from converted speech representations (inference). (c) Prior mixup for better speaker adaptation quality. To reduce the train-inference mismatch problem, the decoder also learns to convert the randomly converted representations into input speech during training.}
\label{fg3}\vspace{-0.5cm}
\end{figure*}

\subsection{Speech Resynthesis}
\label{4-2} 
\vspace{-0.1cm}
\paragraph{Source-filter Encoder}
In this work, we simply define the speech attributes according to the source-filter theory \citep{fant1970acoustic}. The filter encoder takes the content and speaker representations, whereas the source encoder takes the pitch and speaker representations. Previously, \citep{lee2022priorgrad} demonstrated that the data-driven prior in the diffusion process can simply guide the starting point of the reverse process. \citep{popov2022diffusionbased} adopted an average phoneme-level Mel encoder for voice conversion with a data-driven prior. However, this method requires a text transcript to extract the phoneme-level average Mel-spectrogram and pre-trained average Mel-encoder, and the smoothed Mel representation results in mispronunciation. To achieve a substantially more detailed prior, we use the entirely reconstructed source and filter Mel-spectrograms, $Z_{src}$ and $Z_{ftr}$ which are regularized by the target Mel-spectrogram $X_{mel}$ as follows:
\begin{equation}
\label{mel_loss}
  \mathcal{L}_{rec} = \lVert X_{mel} - (Z_{src}+Z_{ftr}) \rVert_1,
\end{equation}
where
\begin{equation}
\label{encoder_out}
 Z_{src} = E_{src}(pitch, s),\;  Z_{ftr} = E_{ftr}(content, s).
\end{equation}
It is worth noting that the disentangled source and filter Mel-spectrograms from the disentangled representations are simply converted with different speaker representation $s$. Thus, we utilize the converted source and filter Mel-spectrogram as each prior in each denoiser for VC. 
\vspace{-0.1cm}
\paragraph{Source-filter Decoder}
We utilize disentangled denoisers for the source and filter representations based on our DDDMs. The source decoder takes a source representation $Z_{src}$ as a prior and the filter decoder takes a filter representation $Z_{ftr}$ as a prior. Subsequently, each denoiser is trained to generate a target Mel-spectrogram from each prior with the same noise, which is conditioned on a speaker representation. Each denoiser can focus on removing the single noise from its own attribute. The forward process is expressed as: 
\begin{equation}
\label{src_fwd}
    dX_{src, t}=\frac{1}{2}\beta_{t}({Z}_{src} - X_{src, t})dt + \sqrt{\beta_{t}}d{W_{t}},
\end{equation}
\begin{equation}
\label{ftr_fwd}
    dX_{ftr, t}=\frac{1}{2}\beta_{t}({Z}_{ftr} - X_{ftr, t})dt + \sqrt{\beta_{t}}d{W_{t}},
\end{equation}
where $t\in [0,1]$, $X_{src,t}$ and $X_{ftr,t}$ are the generated noisy samples with each prior attribute (i.e., source-related and filter-related attribute respectively). For the given forward SDE of each attribute (\ref{src_fwd}) and (\ref{ftr_fwd}), there exist reverse trajectories. The reverse process is expressed as: 
\begin{equation}
\label{src_rev}
\resizebox{.92\linewidth}{!}{$
    d\hat{X}_{src, t} = \left(\frac{1}{2}({Z}_{src} - \hat{X}_{src, t}) -  s_{\theta_{src}}(\hat{X}_{src,t}, {Z}_{src}, s, t) \right.  -  \left. s_{\theta_{ftr}}(\hat{X}_{ftr, t}, {Z}_{ftr}, s, t)
    \right) \beta_{t}dt + \sqrt{\beta_{t}}d\bar{W_{t}},
$}
\end{equation} 
\begin{equation}
\label{ftr_rev}
\resizebox{.92\linewidth}{!}{$
    d\hat{X}_{ftr,t} = \left(\frac{1}{2}({Z}_{ftr} - \hat{X}_{ftr,t}) - s_{\theta_{ftr}}(\hat{X}_{ftr,t}, {Z}_{ftr}, s, t) \right. - \left. s_{\theta_{src}}(\hat{X}_{src,t}, {Z}_{src}, s, t)
    \right)\beta_{t}dt + \sqrt{\beta_{t}}d\bar{W_{t}},
$}
\end{equation}
where $s_{\theta_{src}}$ and $s_{\theta_{ftr}}$ denote the score function parameterized by $\theta_{src}$ and $\theta_{ftr}$ respectively. 

\subsection{Prior Mixup}
\label{4-3}
Although the speech can be disentangled into several attributes and resynthesized with high-quality using the self-supervised representation and diffusion processes, we still train the model by only reconstructing or using the input speech as the target speech in both the reconstruction and diffusion processes, which induces the train-inference mismatch problem. In non-parallel voice conversion scenario, the ground-truth of the converted speech does not exist; Thus, the model is trained only by reconstructing the source speech. However, as we convert the source speech with a different voice style for VC, we shift our focus from reconstruction to conversion even in the training scenario. 

To achieve this, we propose a prior mixup in the diffusion process, which uses the randomly converted representation instead of the reconstructed representation as a prior distribution as illustrated in Figure \ref{fg3-(c)}. Specifically, because the source-filter encoder can also be trained to reconstruct a source and filter of speech from the disentangled representation, the converted source and filter can be obtained with the randomly selected speaker style $s_r$ as follows:
\begin{equation}
\label{conversion_enc}
 Z_{src, r} = E_{src}(pitch, s_{r}), \; Z_{ftr, r} = E_{ftr}(content, s_{r}).
\end{equation}
Subsequently, the randomly converted source and filter, $Z_{src, r}$ and $Z_{ftr, r}$ are used as the prior for each denoiser as below:
\begin{equation}
\label{src_fwd2}
    dX_{src,t}=\frac{1}{2}\beta_{t}({Z}_{src, r} - X_{src,t})dt + \sqrt{\beta_{t}}d{W_{t}}\ ,
\end{equation}
\begin{equation}
\label{ftr_fwd2}
    dX_{ftr,t}=\frac{1}{2}\beta_{t}({Z}_{ftr, r} - X_{ftr,t})dt + \sqrt{\beta_{t}}d{W_{t}}\ 
\end{equation}

The reverse process for the given forward SDE of each attribute (\ref{src_fwd2}) and (\ref{ftr_fwd2}) is expressed as: 
\begin{equation}
\label{src_rev_mixup}
\resizebox{.92\linewidth}{!}{$
    d\hat{X}_{src,t} = \left(\frac{1}{2}({Z}_{src,r} - \hat{X}_{src,t})- s_{\theta_{src}}(\hat{X}_{src,t}, {Z}_{src,r}, {s}, t) \right. -   \left. s_{\theta_{ftr}}(\hat{X}_{ftr,t}, {Z}_{ftr,r}, {s}, t)
    \right)\beta_{t}dt + \sqrt{\beta_{t}}d\bar{W_{t}},
$}
\end{equation} 
\begin{equation}
\label{ftr_rev_mixup}
\resizebox{.92\linewidth}{!}{$
    d\hat{X}_{ftr,t} = \left(\frac{1}{2}({Z}_{ftr,r} - \hat{X}_{ftr,t})- s_{\theta_{ftr}}(\hat{X}_{ftr, t}, {Z}_{ftr,r}, {s}, t) -s_{\theta_{src}}(\hat{X}_{src,t}, {Z}_{src,r}, {s}, t)
    \right)\beta_{t}dt + \sqrt{\beta_{t}}d\bar{W_{t}}.$}
\end{equation}
Hence, the prior mixup can alleviate the train-inference mismatch problem as the model is trained to convert the converted speech into the source speech even when reconstructing the source speech. Moreover, the voice style can be adapted in the source-filter decoder when the source-filter encoder may not execute VC effectively during inference. 
The entire model, including the style encoder, source-filter encoder, and decoder without pre-trained XLS-R and F0 VQ-VAE, is jointly trained in an end-to-end manner with Equation (\ref{dd_rev_train}) for each attribute and Equation (\ref{mel_loss}).

 \paragraph{Training Objectives} As described in section \ref{4-2}, the reconstruction loss $ \mathcal{L}_{rec}$ (\ref{mel_loss}) is used to regulate the encoder output for the data-driven prior of diffusion models.
The reverse SDE of the source attribute (\ref{src_rev_mixup}) and filter attribute (\ref{ftr_rev_mixup}) is trained with the neural network $\theta_{src}$ and $\theta_{ftr}$ to approximate the gradient of the log-density of noisy data $X_{t}$.
Each attribute network is parameterized using the following objective:
\begin{equation}
\label{optim_theta} 
\resizebox{.92\linewidth}{!}{$
\theta^* = \argmin_{\theta}\int_{0}^{1}\lambda_{t}\mathds{E}_{X_{0}, X_{t}}\bigg\Vert\bigg(s_{\theta_{src}}(X_{src,t},Z_{src,r},s, t)+s_{\theta_{ftr}}(X_{ftr,t},Z_{ftr,r},s,t)\bigg)-\nabla{\log{p_{t|0}(X_{t}|X_{0})}}\bigg\Vert_{2}^{2}dt ,
$}
\end{equation}
where $\theta = [\theta_{src}, \theta_{ftr}]$. 
Hence, the diffusion loss can be expressed as the following: 
\begin{equation}
\label{diff_loss_total} 
\resizebox{.92\linewidth}{!}{$
 \mathcal{L}_{diff} = \mathds{E}_{X_{0}, X_{t}}\lambda_{t}\left[\bigg\Vert \bigg(s_{\theta_{src}}(X_{src,t},Z_{src,r},s,t)+s_{\theta_{ftr}}(X_{ftr,t},Z_{ftr,r},s,t) \bigg)-\nabla{\log{p_{t|0}(X_{t}|X_{0})}}\bigg\Vert_{2}^{2}\right] .
$}
\end{equation}
The final objectives for DDDM-VC can be defined as follows:
\begin{equation}
\mathcal{L}_{total} =  \mathcal{L}_{diff} + \lambda_{rec}\mathcal{L}_{rec}\;,
\end{equation} 
where we set $\lambda_{rec}$ to 1. 
\section{Experiment and Result}
\subsection{Experimental Setup}
\paragraph{Datasets}
We used the large-scale multi-speaker LibriTTS dataset to train the model. The \textit{train-clean-360} and \textit{train-clean-100} subsets of LibriTTS, which consist of 110 hours of audio samples for 1,151 speakers, were used for training. Thereafter, we evaluated VC performance on LibriTTS and VCTK dataset for many-to-many and zero-shot VC scenarios. For zero-shot cross-lingual voice conversion scenarios, we used the CSS10 dataset which includes 10 different languages. 

\paragraph{Preprocessing}
We resampled the audio from the sampling rate of 24,000 Hz to 16,000 Hz using the Kaiser-best algorithm of torchaudio Python package. We use the downsampled audio waveform as the input for XLS-R (0.3B) \citep{babu22_interspeech} to extract the self-supervised speech representation. For the target speech and the input of speaker encoder, we used log-scale Mel-spectrogram with 80 bins. To map the time frames between the self-supervised representation and Mel-spectrogram without any interpolation, Mel-spectrogram was transformed with hop size of 320, window size of 1280, and 1280-point Fourier transform.    

\paragraph{Training}
For reproducibility, we attached the source code of DDDM-VC in the Supplementary materials. We trained DDDM-VC using the AdamW optimizer \citep{loshchilov2018decoupled} with $\beta_1=0.8$, $\beta_2=0.99$, and weight decay $\lambda=0.01$, and applied the learning rate schedule with a decay of $0.999^{1/8}$ at an initial learning rate of $5\times10^{-5}$. We train all models including ablation study with a batch size of 64 for 200 epochs. Architecture details are described in Appendix A. For prior mixup, we mixed the speaker representation using binary selection between the original and shuffled representations in the same batch. For zero-shot voice conversion, we did not fine-tune the model. For one-shot speaker adaptation, we fine-tuned the model with only one sentence of novel speakers for 500 steps with optimizer initialization and an initial learning rate of $2\times10^{-5}$. We used the pre-trained Vocoder to convert the Mel-spectrogram into waveform. For vocoder, we used HiFi-GAN V1 \citep{kong2020hifi} as an generator, and we used multi-scale STFT-based discriminators (MS-STFTD) of EnCodec \citep{defossez2022high} which use a complex-valued STFT with real and imaginary components. 

\subsection{Evaluation Metrics}
\paragraph{Subjective Metrics}
We measured the mean opinion score (MOS) for the speech naturalness and speaker similarity in VC tasks. At least 20 listeners rated each sample from the source and converted speech on a scale of 1 to 5 for the speech naturalness MOS (nMOS). At least 20 listeners rated the target and converted speech on a scale of 1 to 4 for the speaker similarity MOS (sMOS).    
\paragraph{Objective Metrics}
We calculated the character error rate (CER) and word error rate (WER) using Whisper \citep{radford2022robust} which is public available automatic speech recognition (ASR) model\footnote{https://github.com/openai/whisper. We used a large model of Whisper with 1,550M parameters, and used a presented text normalizer before calculating the CER and WER.} with large-scale multi-lingual and multitask supervision for the content consistency measurement. We evaluated the equal error rate (EER) of automatic speaker verification (ASV) model \citep{kwon2021ins}, which is trained with large-scale speech recognition dataset, VoxCeleb2 \citep{chung2018voxceleb2} for the speaker similarity measurement. Furthermore, we determined the speaker encoder cosine similarity (SECS) for the additional similarity measurement. As VCTK provided a paired utterance per speaker, we also evaluated the Mel-cepstral distortion (MCD). We produced all possible pairs from the converted and target speech (400$\times$20 = 8,000), and calculated all the evaluation metrics. 
\begin{table*}[t]
  \centering
    \caption{Many-to-many VC results on seen speakers from LibriTTS dataset} 
  \label{table1}
      \resizebox{1\textwidth}{!}{
  \begin{tabular}{l|c|cc|cc|cc|c}
    \toprule
     Method &iter.&  nMOS ($\uparrow$) & sMOS ($\uparrow$)  & CER ($\downarrow$)  & WER ($\downarrow$) & EER ($\downarrow$) &SECS ($\uparrow$)& Params. ($\downarrow$)  \\
    \midrule
     GT  &-&  3.82$\pm$0.05 & 3.44$\pm$0.03& 0.54 & 1.84 & - &-&- \\
      GT (Mel + Vocoder) & - & 3.81$\pm$0.05 & 3.23$\pm$0.05 & 0.60& 2.19 & - &0.986&13M \\
            \midrule
      AutoVC \citep{qian2019autovc} & - & 3.62$\pm$0.05 & 2.44$\pm$0.04& 5.34 & 8.53 & 33.30 &0.703& 30M\\
      VoiceMixer \citep{lee2021voicemixer} & - & 3.75$\pm$0.05 & 2.74$\pm$0.05& 2.39 & 4.20 & 16.00 &0.779&52M   \\
      SR \citep{polyak21_interspeech} & - & 3.62$\pm$0.05 & 2.55$\pm$0.04& 6.63 & 11.72 & 33.30 & 0.693
 &15M \\
         \midrule
      DiffVC \citep{popov2022diffusionbased}&6 &  3.77$\pm$0.05 &2.72$\pm$0.05 & 7.28& 12.80 & 10.50&0.817 &123M \\
      DiffVC \citep{popov2022diffusionbased}&30 &  3.77$\pm$0.05 & 2.77$\pm$0.05 & 7.99 & 13.92 & 11.00 &0.817&123M \\
      DDDM-VC-Small (Ours) &6&  3.75$\pm$0.05&2.75$\pm$0.05 & 3.25& 5.80&6.25 &0.826&21M\\
    DDDM-VC-Small (Ours) &30& $\textbf{3.79}$$\pm$ $\textbf{0.05}$ & $ \textbf{2.81}$ $\pm$ $\textbf{0.05}$ & 4.25& 7.51&6.25 &0.827&21M  \\
      DDDM-VC-Base (Ours) &6&  3.75$\pm$0.05 & 2.75$\pm$0.05& $\textbf{1.75}$ & $\textbf{4.09}$ &$\textbf{4.00}$ &0.843 &66M\\
    DDDM-VC-Base (Ours) &30&  $\textbf{3.79}$$ \pm$$ \textbf{0.05}$  & 2.80$\pm$0.05 &2.60& 5.32& 4.24& $ \textbf{0.845}$  &66M\\
    \bottomrule
  \end{tabular}
}
\end{table*}
\begin{table*}[t] 
\caption{Zero-shot VC results on unseen speakers from VCTK dataset. We additionally report the  one-shot speaker adaptation result of DDDM-VC model (DDDM-VC-Fine-tuing) which is fine-tuned with only single sample per speaker for 500 steps.}
  \centering 
 \resizebox{1\textwidth}{!}{
  \begin{tabular}{l|c|cc|cc|ccc}
    \toprule
     Method &iter.&  nMOS ($\uparrow$) & sMOS ($\uparrow$)  & CER ($\downarrow$)  & WER ($\downarrow$) & EER ($\downarrow$) &SECS ($\uparrow$)& MCD$_{13 }$ ($\downarrow$)   \\
    \midrule
     GT  & - &  4.28$\pm$0.06 & 3.87$\pm$0.03 & 0.21 & 2.17 & - & - & - \\
     GT (Mel + Vocoder) & - & 4.03$\pm$0.07 & 3.82$\pm$0.03 & 0.21 & 2.17 & - & 0.989 & 0.67\\
            \midrule
      AutoVC \citep{qian2019autovc} & - & 2.49$\pm$0.09 & 1.88$\pm$0.08 &5.14 & 10.55& 37.32 &0.715 & 5.01 \\
      VoiceMixer \citep{lee2021voicemixer} & - &  3.43$\pm$0.08 & 2.63$\pm$0.08 & 1.08 & 3.31 & 20.75&0.797& 4.49   \\
      SR \citep{polyak21_interspeech} & - & 2.58$\pm$0.10 & 2.03$\pm$0.07 & 2.12 & 6.18 & 27.24 &0.750 & 5.12 \\
         \midrule
      DiffVC \citep{popov2022diffusionbased}&6 &  3.48$\pm$0.07 & 2.62$\pm$0.08 & 5.82 & 11.76 & 25.30 & 0.786&4.82 \\
      DiffVC \citep{popov2022diffusionbased}&30 &  3.62$\pm$0.07 & 2.50$\pm$0.07 & 6.92 & 13.19 & 24.01 & 0.785& 5.00  \\
    DDDM-VC-Small (Ours) &6&  3.76$\pm$0.07 & 2.99$\pm$0.07 & 1.27 & 3.77 & 6.51&0.852 &\textbf{4.39}\\
    DDDM-VC-Small (Ours) &30&  3.84$\pm$0.06 & 2.96$\pm$0.07 & 1.95 & 4.70 & 6.89 &0.851&4.55\\
    DDDM-VC (Ours) &6&   3.74$\pm$0.07 & 2.98$\pm$0.07 & $\textbf{1.00}$ & $\textbf{3.49}$ & $\textbf{6.25}$&0.856&4.42 \\
    DDDM-VC (Ours) &30&   $\textbf{3.88}$$\pm\textbf{0.06}$ & $\textbf{3.05}$$\pm\textbf{0.07}$ & 1.77& 4.35 & 6.49 &$\textbf{0.858}$& 4.54 \\ \midrule
    DDDM-VC-Fine-tuning (Ours) &6&   3.74$\pm$0.07 & 3.07$\pm$0.07 & 1.26 & 3.80 & 0.81&0.910& 4.27 \\
 DDDM-VC-Fine-tuning (Ours) &30&   3.86$\pm$0.07 & 3.06$\pm$0.07 & 1.87& 4.63 & 0.82 &0.913& 4.38 \\ 
    \bottomrule
  \end{tabular}}
  \label{zvct2} 
\end{table*}
\begin{table*}[t]
  \centering 
    \caption{Results of ablation study on many-to-many VC tasks with seen speakers from LibriTTS.}
  \resizebox{1\textwidth}{!}{
  \begin{tabular}{l|c|cc|cc|cc|c}
    \toprule
     Method &iter.&  nMOS ($\uparrow$) & sMOS ($\uparrow$)  & CER ($\downarrow$)  & WER ($\downarrow$) & EER ($\downarrow$) &SECS ($\uparrow$) & Params. ($\downarrow$)   \\
    \midrule
DDDM-VC (Ours) &30&   3.76$\pm$0.05 & 3.08$\pm$0.05 & 2.60& 5.32 & 4.24 & 0.845 &66M \\
    w.o Prior Mixup &30&  3.79$\pm$0.05 & 3.03$\pm$0.05 & 3.28& 5.66&7.99 &0.821 &66M \\
    w.o Disentangled Denoiser &30&  3.76$\pm$0.05 & 3.00$\pm$0.05 & 3.20& 5.57&9.75&0.815&36M\\
    w.o Normalized F0 &30&  3.78$\pm$0.05& 3.00$\pm$0.05 & 3.27&5.88& 10.25&0.811 &33M\\
    w.o Data-driven Prior &30&  3.83$\pm$0.05 & 2.87$\pm$0.05 & 2.32& 4.86&19.25&0.786 &66M\\
    \bottomrule
  \end{tabular}}
\label{ablationT}
\end{table*}
\subsection{Many-to-Many Voice Conversion}
\label{mmvc}
We performed the many-to-many VC task with seen speakers during the training, and compared our models with various VC models. As indicated in Table \ref{table1}, DDDM-VC-Small also outperformed the other models in all subjective and objective metrics without ASR results. Although VoiceMixer had a lower CER and WER, it had a lower voice style transfer performance in terms of the EER and SECS. Furthermore, we compared the converted speech generated with 6 and 30 iterations to evaluate the performance with fast sampling. Although the objective results of the model with 6 iterations were better than those of the model with 30 iterations, the model with 30 iterations achieved better performance in both the nMOS and sMOS evaluations. Thus, the audio quality was perceptually improved and the generated samples had better diversity with the stochastic iterative processes. 
\vspace{-0.1cm}
\subsection{Zero-shot Voice Conversion}
\label{zvc}
We also report the results of the zero-shot VC tasks. As indicated in Table \ref{zvct2}, our models significantly outperformed the baseline models in terms of speaker similarity. In particular, only the DDDM-VC models could adapt the voice style with novel speakers in terms of EER and SECS. We found that increasing iteration steps improved the diversity of converted speech in that CER, WER, and EER were increased, but the nMOS was consistently improved. We analyzed the effectiveness of each proposed component in the ablation study. In addition, we can control each attribute by transferring different styles to each attribute respectively as indicated in Appendix E.   

\subsection{One-shot Speaker Adaptation}
\label{onevc}
For better speaker adaptation, we additionally fine-tuned our model on the VCTK dataset. We only used one sample per speaker, which is under ten seconds per speaker. As indicated in Table \ref{zvct2}, the speaker similarity in terms of EER and SECS is consistently improved but the CER increased after the model overfitted the small training samples. With a small iteration of training, our model trained with large-scale speech dataset could effectively adapt to novel speaker by only one sample.     

\begin{wrapfigure}{r}{0.45\columnwidth}
  \centering\vspace{-1.15cm}
     \resizebox{0.45\columnwidth}{!}{{\includegraphics[width=1\columnwidth]{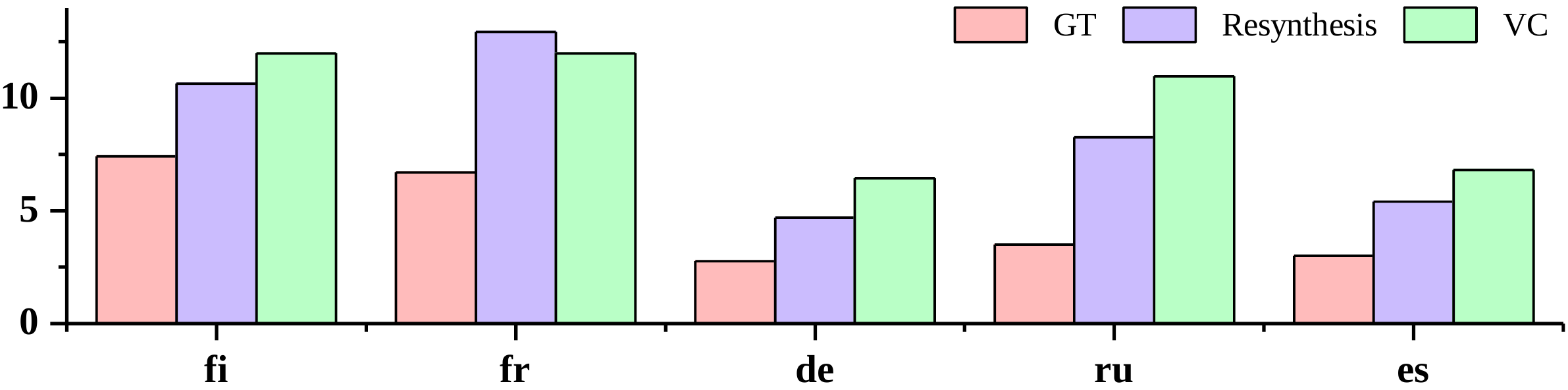}}
     }\vspace{-0.1cm}
\caption{CER results for zero-shot cross-lingual VC on unseen languages from CSS10 multi-lingual dataset.}
  \label{fg4} \vspace{-0.35cm}
\end{wrapfigure}

\subsection{Zero-shot Cross-lingual Voice Conversion}
We performed the zero-shot cross-lingual VC to demonstrate the zero-shot generation performance, even with unseen languages. We first produced all possible pairs from two samples of each language (20$\times$20=400). Subsequently, we calculated the EER of all speakers of all languages (400$\times$20=8000), the results reveal an EER of 9.75$\%$ which is similar to the zero-shot performance of seen language. Moreover, the CER results in Figure \ref{fg4} demonstrate that our model could perform generalization for disentangling and converting speech even in zero-shot cross-lingual scenarios. 
\subsection{Ablation Study}
\label{5-6}    
\paragraph{Prior Mixup} We trained the DDDM-VC model without the prior mixup to clarify the reduction in the train-inference mismatch. 
As indicated in Table \ref{ablationT}, the prior mixup could improve the generalization performance with better speaker adaptation in that the EER of the model with the prior mixup decreased and the SECS increased. However, the naturalness was slightly decreased, which can occur in VC since it does not take into account the target rhythm on the fixed-length of input speech. The research on the rhythm conversion could address this issue and we leave it for the future work.   
\paragraph{Disentangled Denoiser} We observed that removing the disentangled denoiser (employing only a single denoiser) decreased the performance in all metrics. It indicates that the disentangled denoiser can improve the model performance by effectively adapting each representation to the target voice style, compared to a single denoiser.  
\paragraph{Normalized F0} We determined that removing the normalized F0 conditioning decreases the VC performance. Without the pitch contour, the encoder may not disentangle the content information of the speech effectively, resulting in a degradation of the VC performance. As it is difficult to reconstruct the speech from the perturbed speech representation, the use of additional pitch information that can be extracted from the ground-truth speech may improve the stability of the model.     
\paragraph{Data-driven Prior} As noted in \citep{lee2022priorgrad}, a data-driven prior can improve the performance of diffusion model. We minimize the L1 distance of Mel-spectrogram between the ground-truth Mel-spectrogram and output of the source-filter encoder as Equation (\ref{encoder_out}) for the data-driven prior. Each output from the source and filter encoder was used for the prior of each diffusion model, which was disentangled by the source-filter theory. Although nMOS was reported slightly lower, the performance of speaker adaptation significantly increased with data-driven prior. In the VC tasks, using the converted Mel-spectrogram performs better than using the average Mel-spectrogram \citep{popov2022diffusionbased}. Besides, we think that the enhanced prior through normalizing flow \citep{kim2020glow, ren2021portaspeech} may also improve the performance of models.
\section{Conclusion}\vspace{-0.17cm}
We have presented DDDMs for the robust control of various data components in diffusion models. We successfully demonstrated that DDDMs can improve the style transfer performance in VC tasks. DDDM-VC can convert the voice style even in zero-shot voice style transfer tasks by improving the speaker adaptation quality significantly. We have also proposed the prior mixup, which can improve the robustness of style control by learning to restore the data from converted representations for better generalization with reduced train-inference mismatch. Furthermore, we demonstrated that our model can robustly convert the voice with high-quality regardless of the model size. The small model also achieved better performance than state-of-the-art VC models.  
\section{Broader Impact and Limitation}
\label{broader}
\paragraph{Practical Application}
We present DDDMs, which can control the style for each attribute in generative models. We verify the effectiveness of DDDMs with DDDM-VC which can convert the voice style by disentangling the speech and resynthesizing the speech from the disentangled representation. These VC systems could be utilized in various applications such as dubbing systems for the game and film industries. For more practical application, we also extend our model to text-to-speech system by utilizing the pre-trained DDDM-VC in Appendix \ref{section_tts}. In addition, we also present an audio mixing system, DDDM-Mixer in Appendix \ref{section_mixing}.    
\paragraph{Social Negative Impact}
Although TTS or VC systems could be positively utilized in various applications, they also have possible threats of malicious uses such as fake audio generation and voice spoofing. To alleviate these potential harms, audio fingerprint and fake audio detection systems should be presented with the development of speech technology. 
\paragraph{Limitation} Although our model can improve the speaker adaptation quality significantly, there is room for improvement in the speech naturalness for zero-shot cross-lingual VC or noisy speech scenarios, which results from the inaccurate pitch modeling and style transfer with noise. Hence, in future works, we will first attempt to train the model with a cross-lingual dataset using the language-independent speech decomposition to improve the speech naturalness. In addition, we will separate the noise from speech for noise-free voice conversion with noise disentanglement.


\section*{Acknowledgements}
This work was supported by Institute of Information \& communications Technology Planning \& Evaluation (IITP) grant funded by the Korea government(MSIT) (No. 2019-0-00079, Artificial Intelligence Graduate School Program(Korea University), No. 2019-0-01371, Development of Brain-inspired AI with Human-like Intelligence, No. 2021-0-02068, Artificial Intelligence Innovation Hub, and No. 2022-0-00984, Development of Artificial Intelligence Technology for Personalized Plug-and-Play Explanation and Verification of Explanation) and ESTsoft Corp., Seoul, Korea.
\bibliographystyle{plainnat}
\bibliography{biblio}



\newpage
\appendix

\onecolumn
\begin{table*}[h]
  \caption{Hyperparameters of DDDM-VC.}
  \label{hyper}
  \centering
      \resizebox{1\textwidth}{!}{
  \begin{tabular}{c|l|c}
    \toprule
    &Hyperparameter          & DDDM-VC-Small / DDDM-VC   \\
    \midrule
    & Source input  & VQ Codebook of normalized F0\\
    & Source unit size & 20\\
    & Source embedding (Projection) & 128\\
  Source Encoder    & WaveNet Layers & 8\\
    & WaveNet  Kernel & 3\\
    & WaveNet Conv1D Filter Size & 128 \\
    &  Output Size & 128\\
        \midrule
    & Filter input  & 12th layer of XLS-R\\
    & Filter input size & 1024\\
    & Filter embedding (Projection) & 128\\
  Filter Encoder    & WaveNet Layers & 8\\
    & WaveNet  Kernel & 3\\
    & WaveNet Conv1D Filter Size & 128 \\
    &  Output Size & 128\\
     \midrule
     & Input Feature & Mel-spectrogram \\
    & Input Size& 80\\
  & Spectral Encoder Layers & 2\\
   & Spectral Encoder Conv1D Kernel & 5\\
  Style Encoder  & Temporal Encoder Layers  & 2\\
 & Temporal Encoder Conv1D Kernel & 5\\
    & Hidden Size& 256\\
    & Multi-head Attention Head &2 \\
    & Dropout & 0.1\\
     \midrule
     & 2D UNet Layer& 3\\
     & 2D UNet initial size & 64/128 \\
    & 2D UNet up/down size & [1,2,4]\\
 Source-filter Diffusion     & Style conditioning size & 128\\
    &  Source/Filter conditioning size & 80\\
     & $\beta$ Min & 0.05\\
     & $\beta$ Max & 20.0\\
      \midrule
      &Batch Size & 64\\
   Training  & Initial Learning Rate & 5$\times$10$^{-5}$\\
         &audio segmentation & 35840\\
   & $\lambda_{rec}$ &1 \\

    \bottomrule
  \end{tabular}
  }
\end{table*}

\appendix
\section{Implementation Details}
\subsection{DDDM-VC}
 The hyperparameters of DDDM-VC are described in Table \ref{hyper}. DDDM-VC-Small has all the same hyperparameters without the initial dimension size of the source-filter diffusion decoder. For efficient training, we segment audio into 35,840 frames per audio, which is downsampled to 112 frames by XLS-R and this size is same as the Mel-spectrogram size transformed by the hop size of 320. We train all models with the batch size of 64 on two NVIDIA A100 GPUs (batch size of 32 per GPU) for three days. For the F0 VQVAE model, we utilize the official source code of speech resynthesis\footnote{\url{https://github.com/facebookresearch/speech-resynthesis}}. We use the same dataset (\textit{train-clean-360} and \textit{train-clean-100} of LibriTTS subsets), and we train the F0 VQVAE module with the batch size of 16 on single NVIDIA A100 GPUs for 720k steps (It takes only eight hours). For XLS-R (0.3B), we utilize the pre-trained XLS-R\footnote{\url{https://huggingface.co/docs/transformers/model\_doc/xls\_r}}.

\subsection{Baseline Models}
We compared our model with AutoVC \citep{qian2019autovc}, VoiceMixer \citep{lee2021voicemixer}, Speech Resynthesis (SR) \citep{polyak21_interspeech}, and DiffVC \citep{popov2022diffusionbased}. All of the models are trained with the same training dataset of \textit{train-clean} and \textit{train-other} of LibriTTS subsets. For AutoVC and VoiceMixer, the Mel-spectrogram segmented by 192 frames is fed to the model. We use an information bottleneck size of 32 for AutoVC\footnote{\url{https://github.com/auspicious3000/autovc}}. For VoiceMixer, we set the k as 16 (about 0.3 seconds for a sampling rate of 16,000 Hz and hop size of 320). 0.3 seconds means over the average boundary of consonant-vowel syllables \citep{lee2021voicemixer}. For SR, we follow the official source code, however we found that training this model with a large-scale speaker dataset decreased the speaker adaptation performance. Also, although we utilize the same pre-trained F0 VQVAE module that is used to train our model, SR synthesizes some noisy sound when F0 is zero. We have also experienced that the E2E models lack speaker adaptation ability, and the SSL features used in SR contain low acoustic information, which results in mispronunciation and over-smoothing problems in the converted speech. We discovered that expanding SR to the large-scale LibriTTS dataset (rather than VCTK which is utilized in the paper) did not ensure the quality of model despite using the official source code of SR and the same VQ-VAE model our model used. Table 1,2 also demonstrated that the SR has a lower speaker adaptation quality. In addition, the E2E model takes more than 10 days to converge, but our model converges in just two days. Hence, we focused on synthesizing a high-fidelity Mel-spectrogram.  For DiffVC, we follow the official source code\footnote{\url{https://github.com/huawei-noah/Speech-Backbones/tree/main/DiffVC}} to train the model. For better speaker adaptation performance of DiffVC, we additionally use the presented \textit{wodyn} setting for speaker conditioning, which uses speaker embedding with additional noise Mel-spectrogram of the target speaker at time $t$.     

\begin{table*}[ht]\vspace{0.3cm}
\centering
  \caption{The vocoder evaluation results. We used the pre-trained HiFi-GAN as a vocoder, and we replaced the multi-scale discriminator and multi-period discriminator with MS-STFTD.}
\resizebox{1\textwidth}{!}{
  \begin{tabular}{l|ccccccc}
    \toprule
     Method & Mel L1 ($\downarrow$)& PESQ$_{wb}$ ($\uparrow$)&   PESQ$_{nb}$ ($\uparrow$) & Pitch ($\downarrow$) & Periodicity ($\downarrow$) & V/UV F1 ($\uparrow$)    \\
    \midrule
     HiFi-GAN  & 0.176 & 3.515   &3.862&20.350& 0.0937 & 0.9599\\
     HiFI-GAN + MS-STFTD &0.130 & 3.560   &3.881&18.667& 0.0882 & 0.9629\\
    \bottomrule
  \end{tabular}
  }    

  \label{vocoder}
 \end{table*}

\subsection{Vocoder} 
To transform Mel-spectrogram into waveform audio, we use the same neural vocoder, HiFi-GAN V1 \citep{kong2020hifi} for all models without speech resynthesis which is an end-to-end waveform synthesis model. However, as the performance of vocoder is the upper-bound of VC models, we would like to improve the quality of vocoder. Hence, we replace the multi-scale discriminator and multi-period discriminator with multi-scale STFT-based discriminators (MS-STFTD) of EnCodec \citep{defossez2022high} which use the real and imaginary components transformed from audio by STFT. Table \ref{vocoder} shows that only replacing the discriminator can improve the quality of vocoder\footnote{For PESQ, we used an open-source implementation from \url{https://github.com/ludlows/PESQ}.\\
For Pitch, Periodicity, and V/UV F1, we follow the implementation from \url{https://github.com/descriptinc/cargan}}, so we used the HiFi-GAN with MS-STFTD for waveform audio generation. We trained this vocoder with the \textit{train-clean-360} and \textit{train-clean-100} of LibriTTS, which are the same subsets of the training dataset for VC models.     
 
\newpage 
\begin{table*}[ht]
\centering
    \caption{Speech resynthesis results from dev-clean subset of LibriTTS}
  \label{sr1}
      \resizebox{0.9\textwidth}{!}{
  \begin{tabular}{l|c|cc|cc|ccc}
    \toprule
     Method &iter.&  CER ($\downarrow$)& WER ($\downarrow$)& EER($\downarrow$)&SECS ($\uparrow$)& Recon. ($\downarrow$)& MCD$_{13}$ ($\downarrow$)& RMSE$_{f0}$ ($\downarrow$)    \\
    \midrule
     GT  &- &3.53& 5.23  &-&-& - & - & -\\
            \midrule
    DiffVC \citep{popov2022diffusionbased}&6 & 6.86& 11.85  &5.06&0.853& 1.25 & 3.78& 31.14\\
    DiffVC \citep{popov2022diffusionbased}&30 & 6.26& 10.47  &8.58&0.857& 1.33 & 4.10 & 32.09\\
    DDDM-VC (Ours) &6&  4.08& 6.10  &0.32&0.917& 0.92 & 2.75&31.26\\
    DDDM-VC (Ours) &30& 3.91& 5.61  &0.89&0.915& 0.99 & 2.99&30.72\\

    \bottomrule
  \end{tabular}
  }

    \vspace{0.5cm}
  \centering
    \caption{Speech resynthesis results from dev-other subset of LibriTTS}
  \label{sr2}
      \resizebox{0.9\textwidth}{!}{
  \begin{tabular}{l|c|cc|cc|ccc}
    \toprule
     Method & iter.& CER ($\downarrow$)& WER ($\downarrow$)&   EER($\downarrow$)&SECS ($\uparrow$)& Recon. ($\downarrow$) & MCD$_{13}$ ($\downarrow$)& RMSE$_{f0}$ ($\downarrow$)    \\
    \midrule
    GT  &- &2.05& 5.46  &-&-& - & -  & -\\
    \midrule
    DiffVC \citep{popov2022diffusionbased}&6 & 11.73& 20.57  &9.09&0.832& 1.26 & 3.91 &29.49\\
    DiffVC \citep{popov2022diffusionbased}&30 &  12.4& 22.38  &9.09&0.835& 1.36 & 4.27&32.44\\
    DDDM-VC (Ours) &6&  6.19& 11.21  &0.75&0.909& 0.92 & 2.90&34.22\\
    DDDM-VC (Ours) &30& 5.82& 12.04  &0.47&0.901& 1.01 &3.15&31.57\\

    \bottomrule
  \end{tabular}
  }    

   \vspace{0.5cm}
  \centering
    \caption{Speech resynthesis results from test-clean subset of LibriTTS}
  \label{sr3}
      \resizebox{0.9\textwidth}{!}{
  \begin{tabular}{l|c|cc|cc|ccc}
    \toprule
     Method &iter.& CER ($\downarrow$)& WER ($\downarrow$)& EER($\downarrow$)&SECS ($\uparrow$)& Recon. ($\downarrow$)& MCD$_{13}$ ($\downarrow$)& RMSE$_{f0}$ ($\downarrow$)    \\
    \midrule
         GT  &- &1.40& 3.84  &-&-& - & - & -\\
            \midrule
    DiffVC \citep{popov2022diffusionbased}&6 & 4.66& 8.87  &7.69&0.858& 1.20 & 3.71 & 32.76\\
    DiffVC \citep{popov2022diffusionbased}&30 &  5.36& 10.19  &10.25&0.868& 1.28 & 3.98 & 31.99\\
    DDDM-VC (Ours) &6&  1.19& 3.24  &0.26&0.929& 0.90 & 2.79&31.55\\
    DDDM-VC (Ours) &30& 2.85& 5.29  &0.53&0.927& 0.98 & 3.03&32.41\\

    \bottomrule
  \end{tabular}
  }  

\vspace{0.5cm}
    \centering
    \caption{Speech resynthesis results from test-other subset of LibriTTS}
  \label{sr4}
      \resizebox{0.9\textwidth}{!}{
  \begin{tabular}{l|c|cc|cc|ccc}
    \toprule
     Method &iter.&  CER ($\downarrow$)& WER ($\downarrow$)& EER($\downarrow$)& SECS ($\uparrow$)&Recon. ($\downarrow$)& MCD$_{13}$ ($\downarrow$)& RMSE$_{f0}$ ($\downarrow$)    \\
    \midrule
     GT  &- &0.95& 2.03  &-&-& - & - & -\\
            \midrule
    DiffVC \citep{popov2022diffusionbased}&6 & 14.56& 23.39  &6.06&0.844& 1.32 & 4.20 &36.76\\
    DiffVC \citep{popov2022diffusionbased}&30 &  17.19& 26.57  &9.09&0.837& 1.38 & 4.50 & 41.28\\
    DDDM-VC (Ours) &6&  5.99& 10.10  &0.09&0.908& 0.98 & 3.06 & 36.70\\
    DDDM-VC (Ours) &30& 4.76& 8.36  &0.47&0.904& 1.05 & 3.35&35.19\\
    \bottomrule
  \end{tabular}
  }

\end{table*}
\section{Evaluation on Speech Resynthesis}
We evaluated the speech resynthesis performance of DDDM-VC to measure the reconstruction performance, and compared it with DiffVC on the all subsets of LibriTTS including \textit{dev-clean}, \textit{dev-other}, \textit{test-clean}, and \textit{test-other}. 
For the speech resynthesis task, we additionally calculate the Mel-spectrogram reconstruction L1 distance (Recon.), Mel-cepstral distortion (MCD$_{13}$), and the root mean square error of F0 (RMSE$_{f0}$) for evaluating reconstruction performance.  
All above subsets are unseen during training, so Table \ref{sr1}, \ref{sr2}, \ref{sr3}, and \ref{sr4} shows the zero-shot speech resynthesis performance. The results indicated that our model outperform the DiffVC in all objective metrics without the RMSE$_{f0}$ of \textit{test-clean subset}. In terms of speaker similarity, DDDM-VC can adapt to novel speaker during speech resynthesis. However, DiffVC, which uses the average Mel-spectrogram as a prior of diffusion model, has a lower speaker adaptation performance even in the speech resynthesis tasks. As shown in Table 3, using the converted Mel-spectrogram as a prior for diffusion models can improve the quality of synthetic speech.

\begin{table*}[t]
  \centering
    \caption{Zero-shot voice conversion results on unseen speakers from LibriTTS-dev-clean subset}
  \label{librivc1}
      \resizebox{0.68\textwidth}{!}{
  \begin{tabular}{l|c|cc|cc}
    \toprule
     Method &iter.&  CER ($\downarrow$)  & WER ($\downarrow$) & EER ($\downarrow$) & SECS ($\uparrow$) \\
    \midrule
     GT  &- & 0.15 & 0.66 & -&-  \\
            \midrule
      DiffVC \citep{popov2022diffusionbased}&6  & 3.77& 7.39 & 20.34 & 0.770 \\
      DiffVC \citep{popov2022diffusionbased}&30  & 4.00 & 7.89 & 22.34& 0.771  \\
    DDDM-VC-Base   (Ours) &6&  2.55& 4.35&6.99 &0.832\\
DDDM-VC-Base  (Ours) &30&  3.26& 5.77 &8.53 &0.833\\
    \bottomrule
  \end{tabular}
  }

\vspace{0.5cm}
  \centering
    \caption{Zero-shot voice conversion results on unseen speakers from LibriTTS-dev-other subset}
  \label{librivc2}
      \resizebox{0.68\textwidth}{!}{
  \begin{tabular}{l|c|cc|cc}
    \toprule
     Method &iter.&  CER ($\downarrow$)  & WER ($\downarrow$) & EER ($\downarrow$) & SECS ($\uparrow$)  \\
    \midrule
     GT  &-&   2.14 & 4.33& - &- \\
            \midrule
      DiffVC \citep{popov2022diffusionbased}&6 & 13.33& 22.81 & 24.09 & 0.750 \\
      DiffVC \citep{popov2022diffusionbased}&30 &  16.56 & 28.97 & 24.50 & 0.757 \\

    DDDM-VC-Base (Ours) &6&   8.23 & 14.93 & 13.67 & 0.818\\
    DDDM-VC-Base (Ours) &30&   11.02 & 20.54 & 12.75 & 0.820\\

    \bottomrule
  \end{tabular}
  }

  \vspace{0.5cm}
  \centering
    \caption{Zero-shot voice conversion results on unseen speakers from LibriTTS-test-clean subset}
  \label{librivc3}
     \resizebox{0.68\textwidth}{!}{
  \begin{tabular}{l|c|cc|cc}
    \toprule
     Method &iter.&  CER ($\downarrow$)  & WER ($\downarrow$) & EER ($\downarrow$) & SECS ($\uparrow$)  \\
    \midrule
     GT  &-&  1.95& 4.00 & - &- \\
            \midrule
      DiffVC \citep{popov2022diffusionbased}&6 & 4.85& 8.50 & 25.33 & 0.768 \\
      DiffVC \citep{popov2022diffusionbased}&30 & 5.57& 9.53 & 25.70 & 0.775  \\

      DDDM-VC-Base (Ours) &6& 6.58 & 8.97 & 9.17 & 0.831\\
      DDDM-VC-Base (Ours) &30&   4.05& 6.87 &10.14 & 0.832\\
    \bottomrule
  \end{tabular}
  }

  \vspace{0.5cm}
    \centering
    \caption{Zero-shot voice conversion results on unseen speakers from LibriTTS-test-other subset}
  \label{librivc4}
    \resizebox{0.68\textwidth}{!}{
  \begin{tabular}{l|c|cc|cc}
    \toprule
     Method &iter.&  CER ($\downarrow$)  & WER ($\downarrow$) & EER ($\downarrow$) & SECS ($\uparrow$)  \\
    \midrule
     GT  &-& 1.44& 3.13 & - &- \\
         \midrule
      DiffVC \citep{popov2022diffusionbased}&6 & 16.66& 26.77 & 20.75 & 0.769 \\
      DiffVC \citep{popov2022diffusionbased}&30 &  20.35& 31.98 & 22.35 & 0.776 \\

      DDDM-VC-Base (Ours) &6& 8.04 & 13.78 & 7.56 & 0.836\\
      DDDM-VC-Base (Ours) &30&   11.50& 18.90 &6.99 & 0.837\\

    \bottomrule
  \end{tabular}
  }

\end{table*}

\newpage
\section{Zero-shot Voice Conversion on LibriTTS}
We additionally evaluate the zero-shot voice conversion performance on the all subsets of LibriTTS including \textit{dev-clean}, \textit{dev-other}, \textit{test-clean}, and \textit{test-other}. The zero-shot voice conversion performance of DDDM-VC outperformed that of DiffVC in all subsets. As Table \ref{librivc2} and \ref{librivc3}, we found that a noisy source and target speech significantly degrade the voice conversion performance in terms of naturalness. We expect the noise modeling in encoder will improve the robustness of voice conversion with better pronunciation and speaker adaptation performance. Specifically, in future work, robust content, pitch, and speaker modeling irrelevant to environmental noise will be conducted for better generalization on the zero-shot voice conversion scenario. 
\newpage

\begin{figure*}[h]
  \centering
    {\includegraphics[width=1\textwidth]{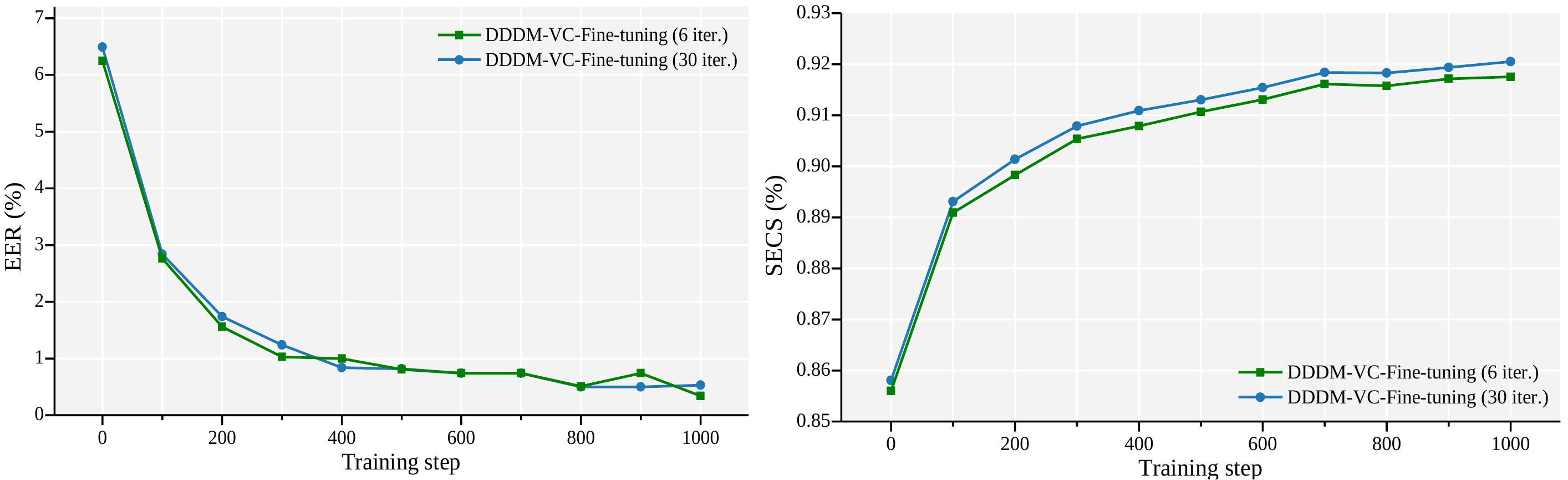}} 
\caption{EER and SECS results according to training steps of fine-tuning for one-shot speaker adaptation.}
  \label{plot1}
\end{figure*}

\begin{figure*}[h]
  \centering
    {\includegraphics[width=1\textwidth]{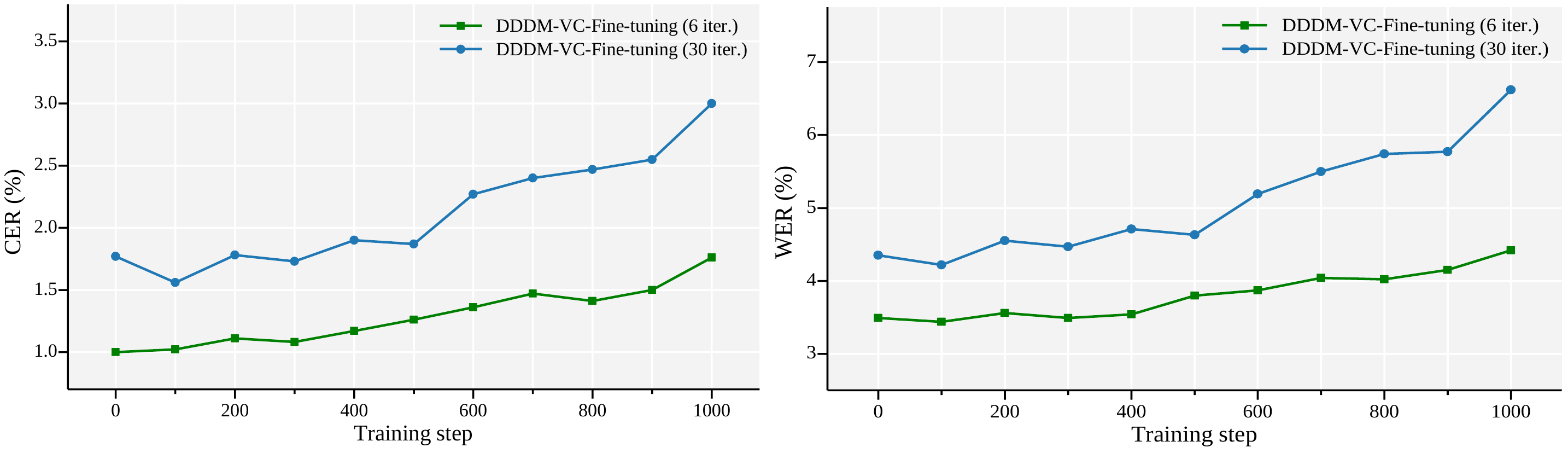}} 
\caption{CER and WER results according to training steps of fine-tuning for one-shot speaker adaptation.}
  \label{plot2} 
\end{figure*}

\section{One-shot Speaker Adaptation}
As indicated in Table \ref{zvct2}, the one-shot speaker adaptation of our model could significantly increase the speaker similarity in terms of EER and SECS. Figure \ref{plot1} also shows the speaker adaptation performance is improved during fine-tuning. However, the CER and WER increased after the model overfits a single sample per speaker. Hence, we only fine-tune the model for 500 steps. Following \citep{kim2022guided2}, we also fine-tune the model with real-world dataset or non-human characters. We highly recommend to listen the audio samples \footnote{\url{https://hayeong0.github.io/DDDM-VC-demo/}}.

\section{Style Control}
\begin{table}[h]
\centering \vspace{-0.5cm}
\caption{Objective evaluation for controlling each attribute. We report the EER and SECS for target speaker as an ground-truth.}\vspace{0.2cm}\label{r1}
\resizebox{0.7\columnwidth}{!}{
\begin{tabular}{l|cc|cc|cc}
\toprule
Method&Filter & Source& CER ($\downarrow$) & WER ($\downarrow$) &  EER ($\downarrow$) &SECS ($\uparrow$)  \\
\midrule
Voice conversion&$s_{trg}$ & $s_{trg}$ & 1.27 & 3.77 &  6.51  & 0.85  \\
Timbre control&$s_{trg}$ & $s_{src}$   & 0.91 & 3.10 & 7.30 & 0.84  \\
Pitch control&$s_{src}$ & $s_{trg}$  & 1.09 & 3.45 & 49.00  & 0.65  \\
Resynthesis&$s_{src}$ & $s_{src}$   & 0.45 & 2.08  & 50.15 & 0.65  \\
\bottomrule
\end{tabular}
} 

\end{table}
Although we can control each attribute by transferring different or mixed voice styles to each attribute respectively, for voice conversion,  we transfer the voice style of the same target speaker to each attribute. Table \ref{r1} reveals that our model can control for each attribute with different styles from source and target speech ($s_{src}$ and $s_{trg}$). Specifically, controlling each filter and source attribute can convert the timbre and pitch, respectively. Although pitch control increases the CER and WER slightly, EER and SECS decrease by converting the intonation with target speaking style.

\newpage

\begin{figure}[h]
  \centering
   \begin{subfigure}[b]{0.45\textwidth}
        \centering
      \includegraphics[height=4.5cm]{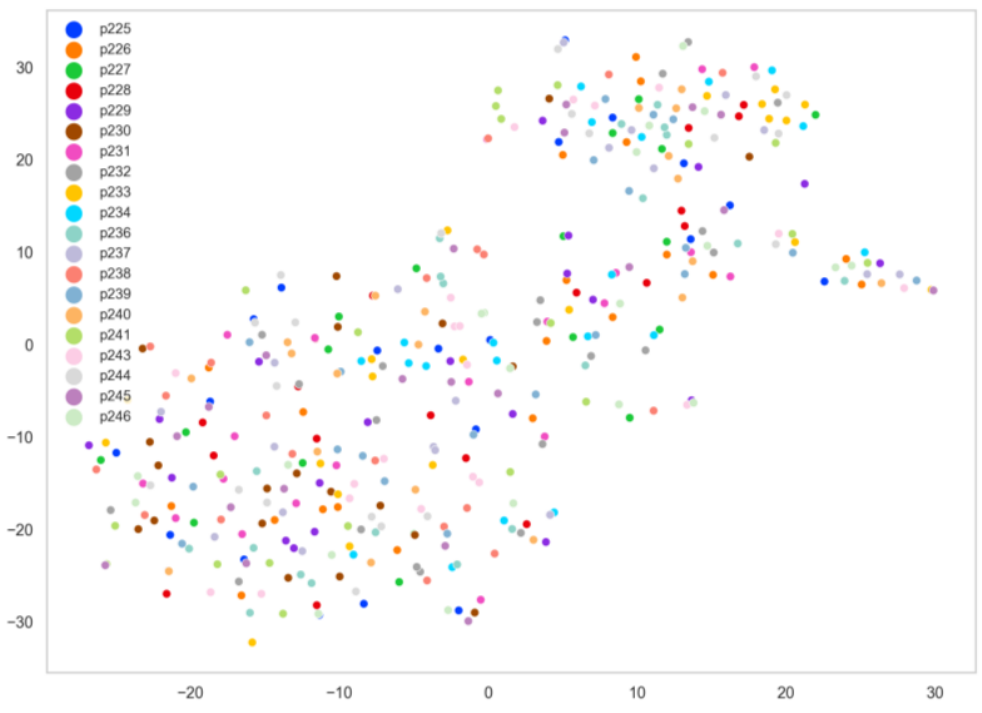}
      \caption{t-SNE visualization of content representation from the ground-truth speech of the VCTK dataset.}\label{f7a}
    \end{subfigure} \\ \vspace{0.3cm}
    \begin{subfigure}[b]{0.45\textwidth}
        \centering
      \includegraphics[width=1\textwidth]{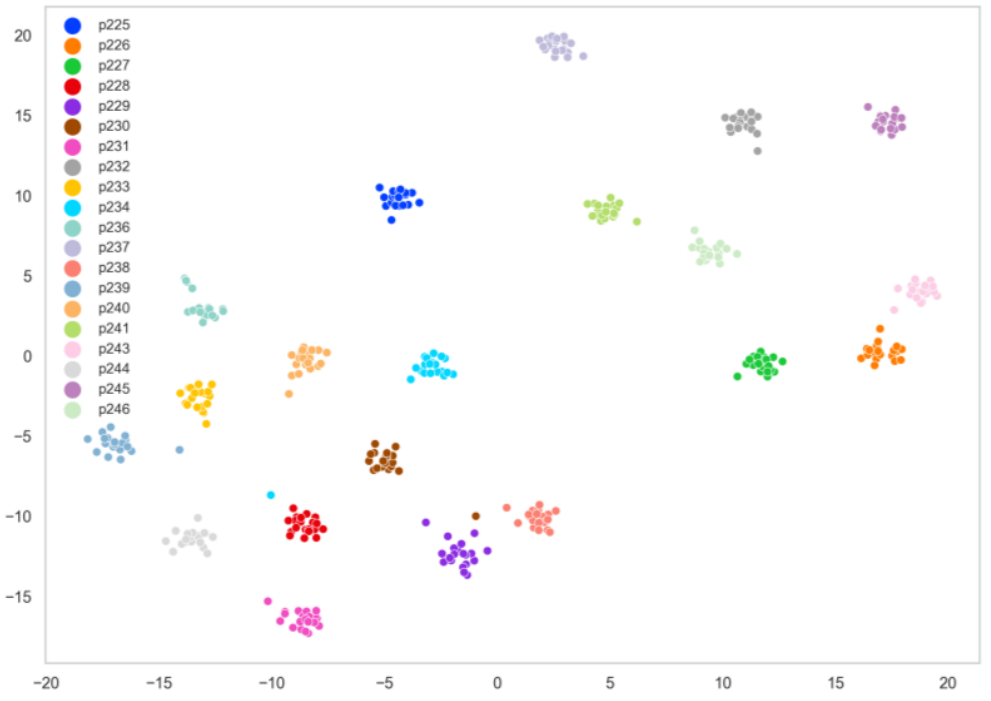}
      \caption{t-SNE visualization of speaker embedding by the Resemblyzer from the ground-truth speech.}\label{f7b}
    \end{subfigure}
     \hspace{0.5cm}
    \begin{subfigure}[b]{0.45\textwidth}
      \centering
      \includegraphics[width=1\textwidth]{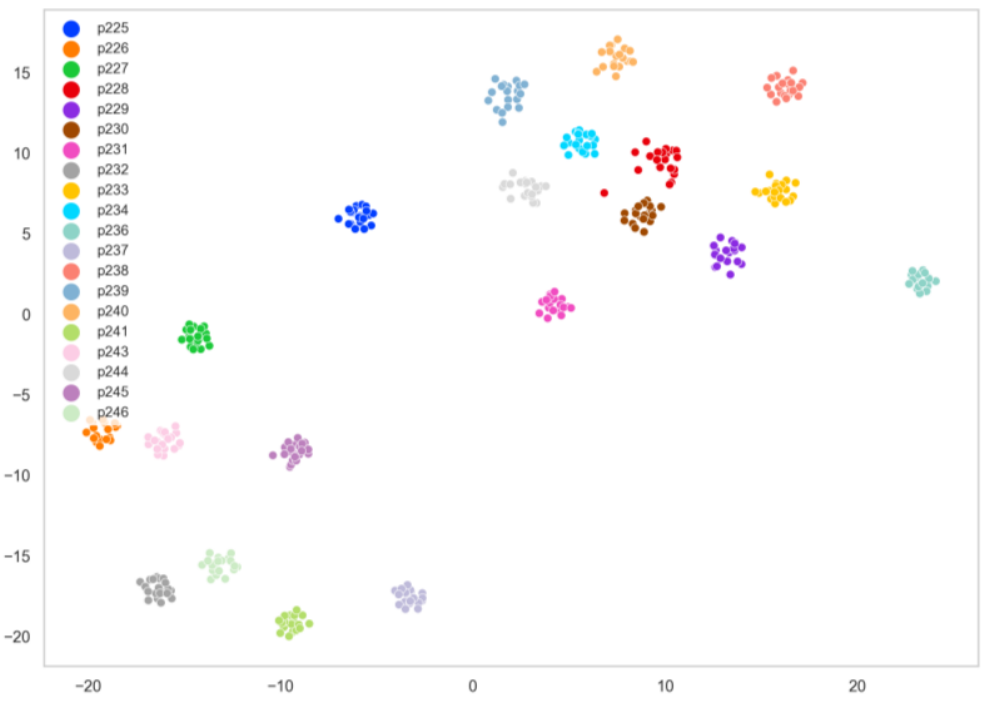}
      \caption{t-SNE visualization of speaker embedding by the Resemblyzer from the converted speech.}\label{f7c}
    \end{subfigure}
  \caption{t-SNE visualization of content and speaker representation.}
\label{fg7}
\end{figure}
\section{t-SNE Visualization}
The t-SNE visualization of content representation, the representation from the 12$^{th}$ layer of XLS-R model, is presented in Figure \ref{f7a}. The content representations for each speaker from VCTK were not clustered by speaker, and it means that the speaker information is disentangled and the content representation only contains the representation irrelevant to speaker information. To demonstrate the zero-shot voice conversion performance, we also present the t-SNE visualization of speaker embedding which is extracted from the ground-truth and converted speech as illustrated in Figure \ref{f7b} and \ref{f7c}. We use the external tool, Resemblyzer, to extract the speaker representation of speech. Both speaker embeddings are clustered by each speaker. This means that our model successfully executes the zero-shot voice conversion.

\newpage
\section{Evaluation Details}

\vspace{-0.1cm}\paragraph{Mean Opinion Score} 

\begin{figure}[h]
  \centering
    \begin{subfigure}[b]{0.45\textwidth}
        \centering
      \includegraphics[height=4.5cm]{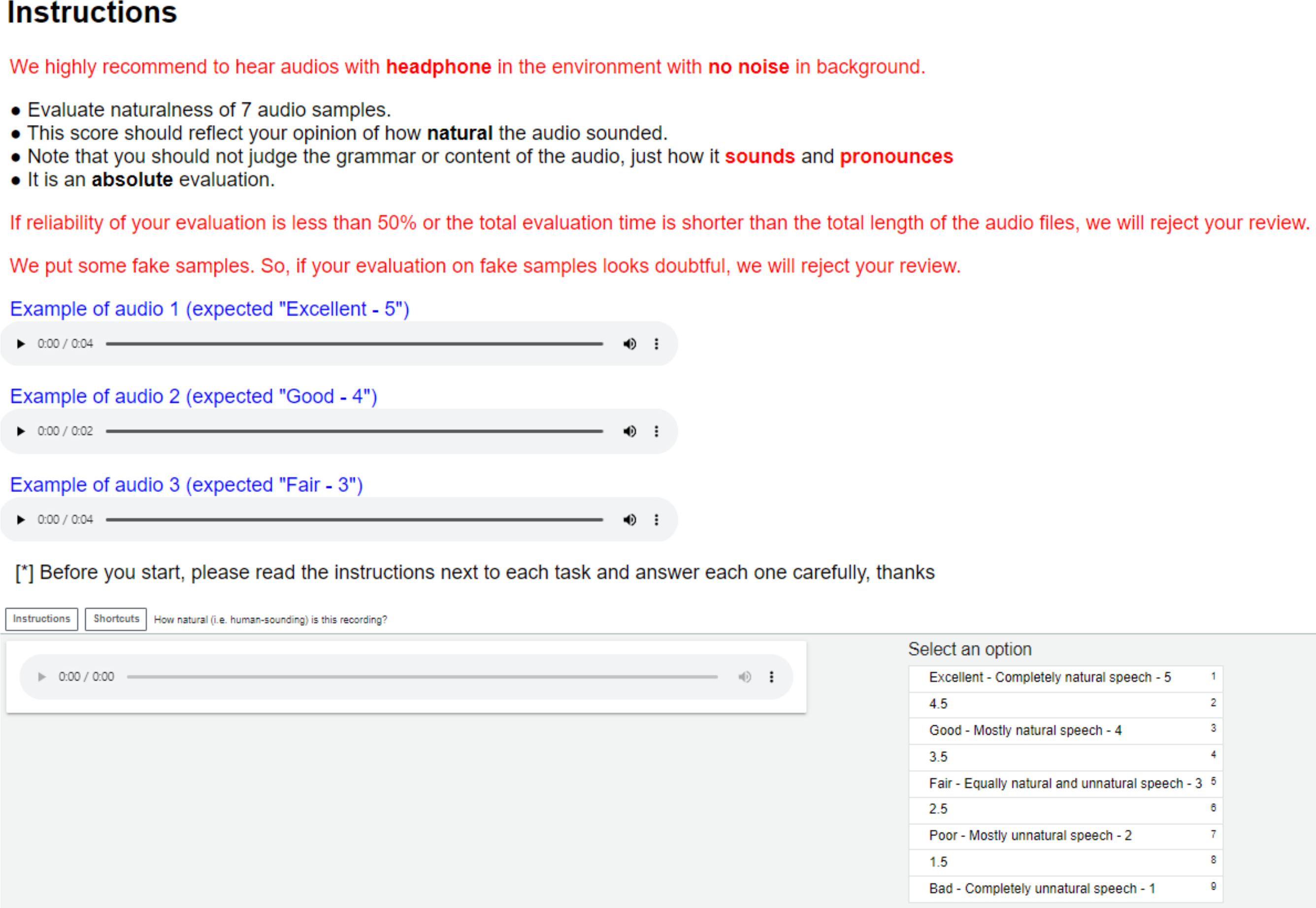}
      \caption{nMOS}
    \end{subfigure} 
    \begin{subfigure}[b]{0.45\textwidth}
      \centering
      \includegraphics[height=4.5cm]{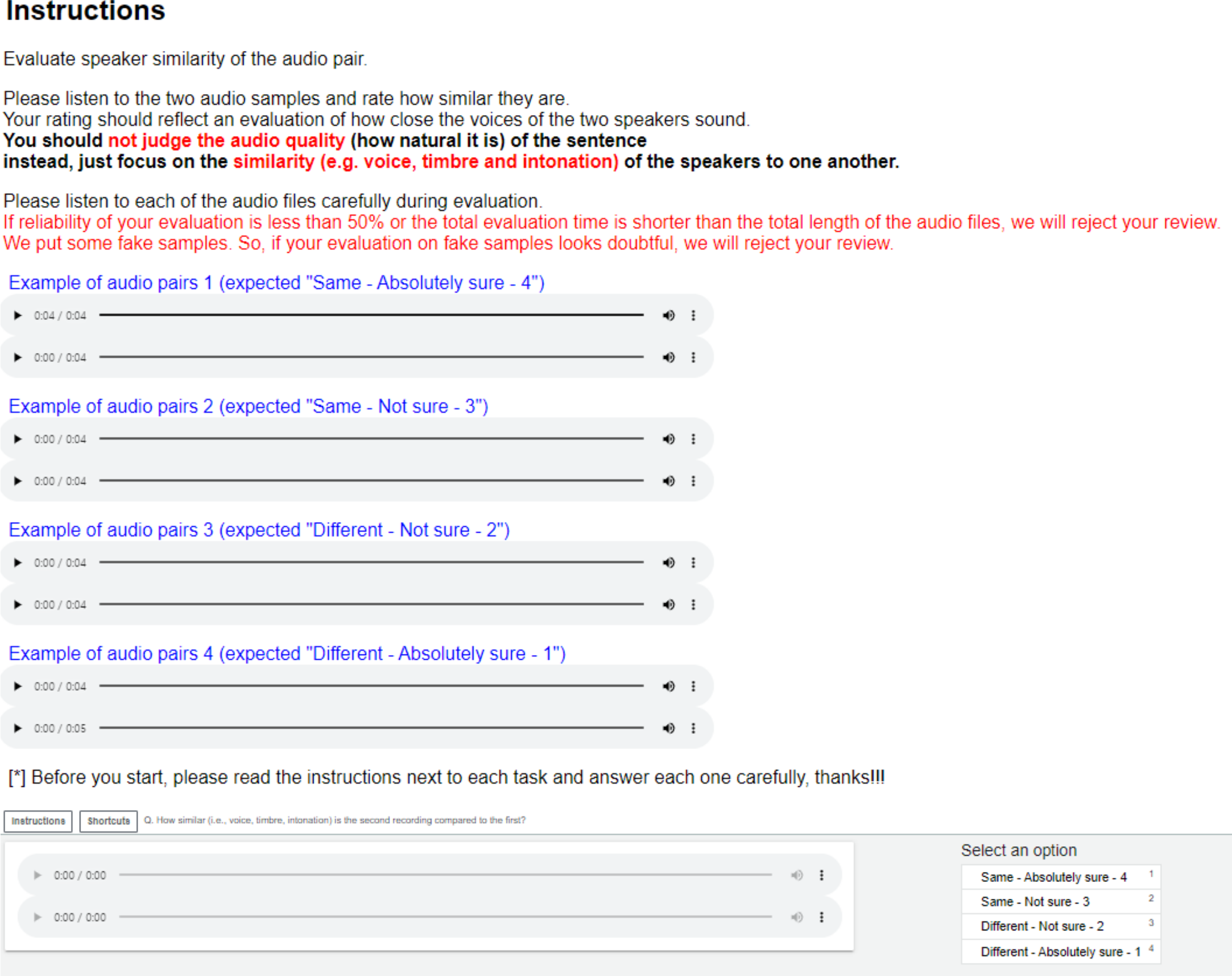}
      \caption{sMOS}
    \end{subfigure}
  \caption{The screenshots of the Amazon MTurk MOS survey. \$0.08 per 1 hit is paid to participants for nMOS and sMOS.}
\label{mos}
\end{figure}

We evaluated MOS to measure the perceptual quality of synthesized speech. The naturalness of a synthesized speech is measured by the nMOS, and the similarity between the synthesized speech and the ground-truth speech is evaluated by the sMOS. We conducted a survey on Amazon Mechanical Turk\footnote{\url{https://www.mturk.com/}} to evaluate the quality of the synthesized speech as shown in Figure \ref{mos}.
\vspace{-0.1cm}\paragraph{Character Error Rate and Word Error Rate}
To evaluate the pronunciation of synthesized speech, we measured character error rate (CER) and word error rate (WER) using Whisper \citep{radford2022robust}\footnote{\url{https://github.com/openai/whisper}}, an automatic speech recognition system trained on multi-lingual and multi-task supervised data.
\vspace{-0.1cm}\paragraph{Equal Error Rate}
The equal error rate (EER) is measured to evaluate the speaker similarity between the ground-truth and  synthesized speech. Specifically, it is useful for quantitatively evaluating the similarity between the converted speech and target speech in the VC task. To this end, we calculated the EER of the automatic speaker verification (ASV) model\footnote{\url{https://github.com/clovaai/voxceleb_trainer}} trained with Voxceleb2  \citep{chung2018voxceleb2}, a large speech recognition dataset containing 5,994 speakers. Specifically, we used the pre-trained model with online augmentation\footnote{\url{https://www.robots.ox.ac.uk/~joon/data/baseline_v2_smproto.model}}, which improves the speaker verification performance of model.
\vspace{-0.1cm}\paragraph{Speaker Encoder Cosine Similarity}
For a more thorough analysis of speaker similarity, we evaluated speaker encoder cosine similarity (SECS). We used Resemblyzer\footnote{\url{https://github.com/resemble-ai/Resemblyzer}}, a speaker encoder capable of high-level feature extraction, to extract speaker features from converted speech and ground-truth speech and then calculated the cosine similarity in the extracted embedding space.
\vspace{-0.1cm}\paragraph{Mel Cepstral Distortion} 
We extracted the first 13 mel-frequency cepstral coefficients (MFCCs) by applying a discrete cosine transform to the ground-truth and synthesized speech\footnote{\url{https://github.com/MTG/essentia/}}. Subsequently, we used the dynamic time warping (DTW) between the extracted MFCCs features, and the sequences of different lengths are aligned to calculate the similarity between the two features.
\begin{equation}
MCD_{13}=\frac{1}{T}\sum_{t=0}^{T-1}{\sqrt{\sum_{k=1}^{13}{(\boldsymbol{M}_{t,k}-\boldsymbol{M}_{t,k}^{'})^{2}}}} ,
\end{equation}
where  $T$ is the number of frames and \(\boldsymbol{M}_{t,k}\), \(\boldsymbol{M}_{t,k}^{'}\) denote the ground-truth and synthesized $k^{th}$ MFCCs of $t^{th}$ frame. We only conduct the MCD evaluation for VCTK dataset, which contains the paired sentences for each speaker.

\section{Audio Mixing}
\label{section_mixing}
\begin{figure*}[t]
    \centering
    {\includegraphics[width=1\textwidth]{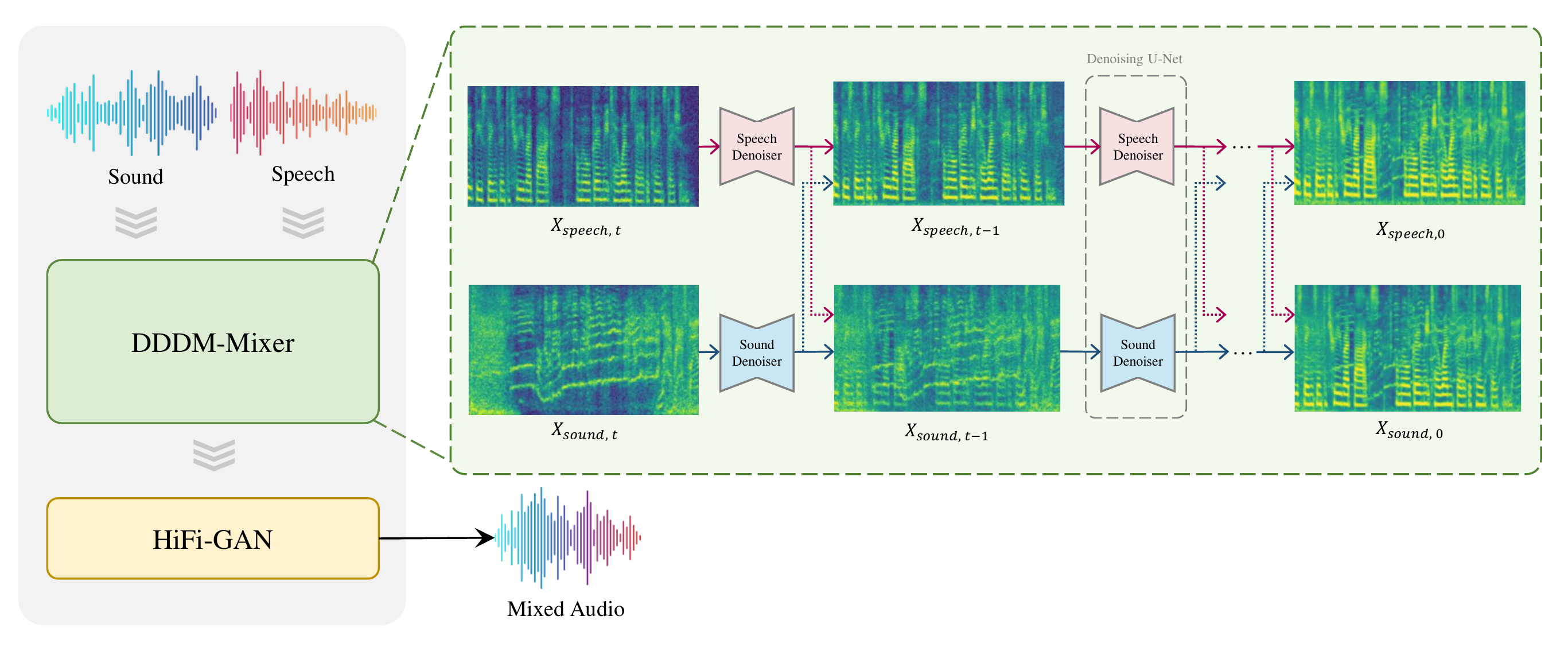}}
    \caption{Overall framework of DDDM-Mixer} 
    \label{dddm-mixer} 
\end{figure*}
\paragraph{DDDM-Mixer} We extend DDDMs to DDDM-Mixer which leverages multiple denoisers to mix the sound and speech into the mixture of Mel-spectrograms by blending them with the desired balance as shown in Figure \ref{dddm-mixer}. 
While DDDM-VC disentangles a single speech into source and filter attributes for attribute denoisers, DDDM-Mixer treats the mixture of audio as a target audio and disentangles it into sound and speech as attributes. We utilize the data augmentation according to signal-to-noise ratio (SNR) from -5 dB to 25 dB for target audio. DDDM-Mixer utilizes the Mel-spectrogram of each attribute as a prior. We employ two denoisers, both of which remove a single noise in terms of their own attribute (sound and speech). We utilize an SNR dB between sound and speech as conditional information to mix each attribute with a desired ratio. We concatenate the SNR positional embedding with the time positional embedding for the condition. Hence, we could mix the generated sound and speech on the Mel-spectrogram. To verify the effectiveness of the DDDM-Mixer, we compare the augmented audio between vocoded sound and speech and the vocoded audio from mixed Mel-spectrogram. The evaluation results show DDDM-Mixer has better performance than audio augmentation according to SNR between the vocoded sound and speech.    
\paragraph{Implementation Details}
We implement the DDDM-Mixer with the same DDDMs as the DDDM-VC-Small, which consists of two 2D UNet-based diffusion models with 3 layers and a hidden size of [64, 128, 256]. We train the DDDM-Mixer utilizing AudioSet, BBC sound effects, Clotho, DEMAND, ESC, FSD18k, Nonspeech100, Sonniss Game Effects, TAU Urban Acoustic Scenes 2019, and UrbanSound8K for sound datasets, and LibriTTS and CSS10 for speech datasets. We utilize the pre-trained HiFi-GAN with speech dataset.     

\begin{table*}[h]
  \centering
    \caption{The objective evaluation of audio mixing }
  \label{tb_mixer}
      \resizebox{0.7\textwidth}{!}{
  \begin{tabular}{l|c|ccc}
    \toprule
     Method & iter.& Mel L1 ($\downarrow$) & PESQ$_{nb}$ ($\uparrow$)  & PESQ$_{wb}$ ($\uparrow$) \\
    \midrule
    Augmentation & -& 0.389 &  2.685 & 3.256 \\
    DDDM-Mixer (Ours) &6& 0.376 & 3.246 &  3.690 \\
    DDDM-Mixer (Ours) &30& 0.335 & 3.264 &  3.708 \\
    \bottomrule
  \end{tabular}
  }
\end{table*}

\paragraph{Evaluation}
We conduct the Mel L1 and PESQ evaluation between the augmented audio and mixed audio using DDDM-Mixer. Table \ref{tb_mixer} shows that DDDM-Mixer has a better audio mixing performance than audio augmentation according to SNR between vocoded sound and speech separately. Figure \ref{dddm-mixer} shows that each sound and speech are mixed by DDDM-Mixer. 

In future works, we will train the HiFi-GAN with both sound and speech datasets, and also utilize the augmented audio for training to convert the mixed Mel-spectrogram into waveform audio robustly.
In addition, we see that we could extend the DDDM-Mixer scenario to mix the sound from the text-to-audio model and the speech from the speech synthesis model.      

\newpage
\begin{figure*}[t]
    \centering
    {\includegraphics[width=1\textwidth]{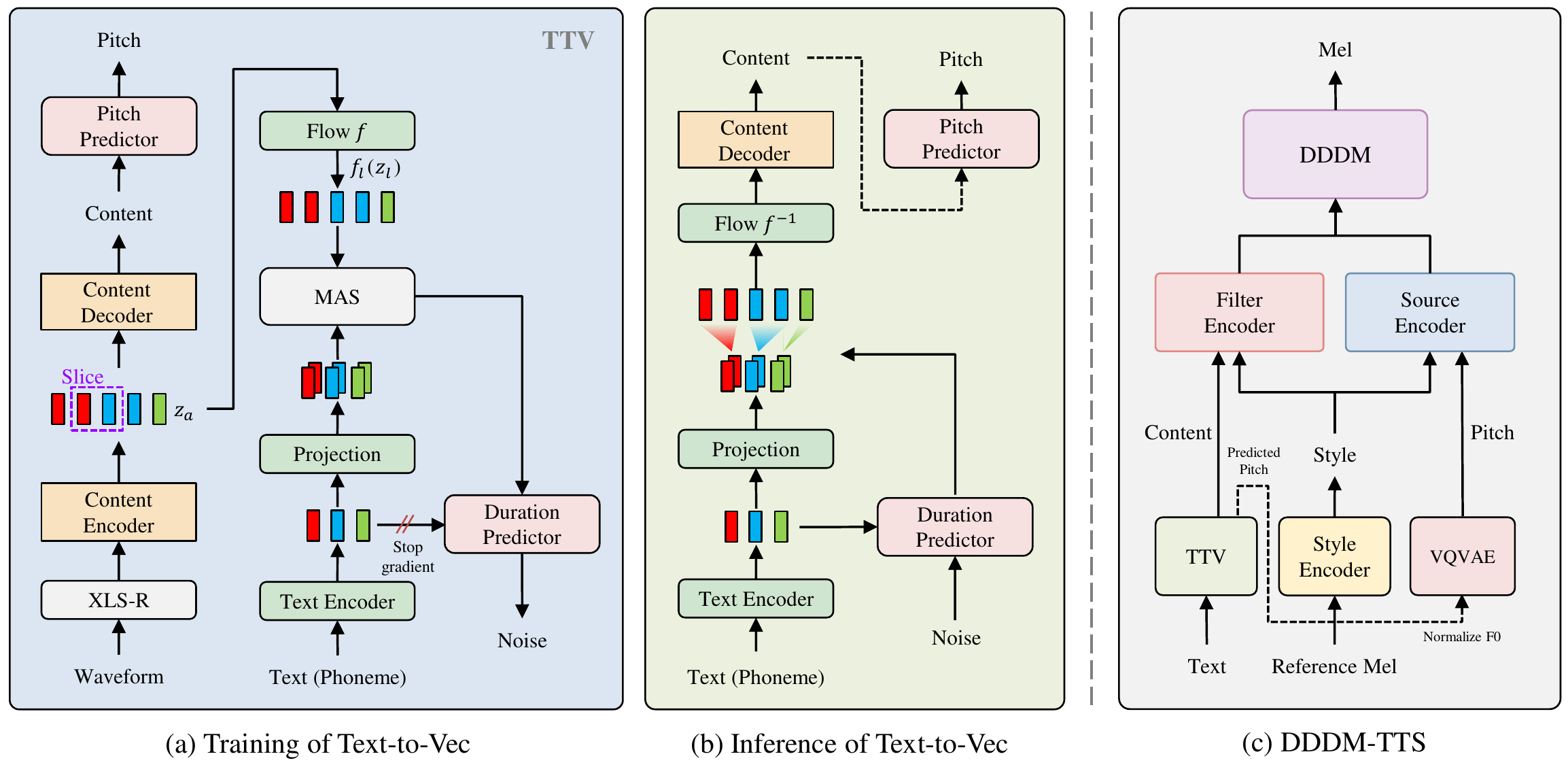}}
    \caption{Overall framework of DDDM-TTS}
    \label{dddm-tts} 
\end{figure*}
\section{Text-to-Speech}
\label{section_tts}
\paragraph{DDDM-TTS} For a practical application, we have experimented with an extension version of DDDM-VC for a text-to-speech system as illustrated in Figure \ref{dddm-tts}. In DDDM-VC, our goal is to train the model without any text transcripts, and only utilize the self-supervised representation for speech disentanglement. Based on the DDDM-VC, we train the text-to-vec (TTV) model which can generate the self-supervised speech representation (the representation from the middle layer of XLS-R) from the text as a content representation. We jointly train the duration predictor and pitch predictor. The predicted content and pitch representation are fed to DDDM-VC instead of each representation from the waveform to synthesize the speech. Hence, we could synthesize the speech from text by utilizing the pre-trained DDDM-VC.   

\paragraph{Text-to-Vec (TTV)} Specifically, we first convert the text prompt into the phoneme sequence. Then, the phoneme sequence is fed to the text encoder to generate the content representation and pitch representation. We utilize the self-supervised speech representation as a content representation and log-scale F0 as a pitch representation. The text encoder consists of a conditional variational autoencoder, duration predictor, and pitch predictor. First, the self-supervised speech representation is encoded and reconstructed by a variational autoencoder. The transformer encoder conditioned by phoneme sequence generates the mean and variance for the linguistic prior distribution, and we utilize monotonic alignment search to align the latent representation from the posterior distribution and the linguistic prior distribution by searching the alignment maximizing the likelihood of data \citep{kim2021conditional}. For better expressiveness of prior distribution, we also use the normalizing flow on the latent representation. In addition, we adopt the stochastic duration predictor of VITS and the pitch predictor consisting of multi-receptive filed fusion module \citep{kong2020hifi} by conditioning voice prompt to predict the speaker-dependent duration and pitch representation from text representation and content representation, respectively. The generated content representation and log-scale F0 from text are fed to the pre-trained DDDM-VC to generate Mel-spectrogram. We called this text-to-speech pipeline DDDM-TTS.    

\paragraph{Implementation Details}
We utilize the Phonemizer\footnote{https://github.com/bootphon/phonemizer} to transform the text sequence into the International Phonetic Alphabet (IPA) sequence. For robust alignment between text and latent representation, we add the blank token between phonemes. Text encoder consists of six blocks of transformer networks which take the sequence of phoneme embedding. Content encoder and decoder consist of 16 layers of Non-casual WaveNet with a hidden size of 256. Flow block consists of four affine coupling layers, and the WaveNet is utilized for the scale and bias of the affine transform. The pitch predictor has two blocks consisting of two upsampling layers and two multi-receptive field fusion (MRF) blocks \citep{kong2020hifi}. We utilize the log-scale F0 for target pitch information, and normalized the predicted F0 before fed to VQVAE. We train the TTV model with a batch size of 128 for 500k steps. Following \citep{kim2021conditional}, we use the AdamW optimizer and two NVIDIA A100 GPUs to train. For the pitch predictor, we use windowed generator training with a randomly sliced F0 of 120 frames.              

\paragraph{Evaluation}We conduct the objective evaluation with text-to-speech models, VITS \citep{kim2021conditional} and HierSpeech \citep{lee2022hierspeech}. For voice style transfer, we utilize the same style encoder for all models. DDDM-TTS has better performance on pronunciation in terms of CER and WER, and also achieves higher speaker adaptation performance in terms of EER and SECS as indicated in Table \ref{ttstable}. We additionally evaluate the zero-shot voice style transfer performance of DDDM-TTS, and Table \ref{ttszero} shows that our model could synthesize the speech from the text even with the novel voice style. We also add the audio samples of DDDM-TTS at the demo page.

We will scale up the DDDM-VC model with a large-scale speech dataset, and use the DDDM-VC as a backbone speech synthesizer for DDDM-TTS. We see that utilizing the pre-trained model with the cross-lingual speech dataset could be extended to the low-resource speech synthesizer from the text for text-to-speech or speech for voice conversion.    
\begin{table*}[t]
  \centering
    \caption{Text-to-Speech results on seen speakers from LibriTTS-train-clean subset}
      \resizebox{0.68\textwidth}{!}{
  \begin{tabular}{l|cc|cc}
    \toprule
     Method &  CER ($\downarrow$)  & WER ($\downarrow$) & EER ($\downarrow$) & SECS ($\uparrow$) \\
    \midrule
       GT & 1.18& 2.56 & - & - \\
       \midrule
       VITS \citep{kim2021conditional}  & 11.75& 19.22 & 5.00 & 0.843 \\
      HierSpeech \citep{lee2022hierspeech} & 6.09 & 9.34 & 11.31& 0.828  \\
    DDDM-TTS  (Ours) &  3.92& 8.13 &3.00 &0.878\\
    \bottomrule
  \end{tabular}
  } 
  \label{ttstable}
\end{table*}
\begin{table*}[t]
  \centering
    \caption{Text-to-Speech results on unseen speakers from VCTK dataset}
  \label{ttszero}
      \resizebox{0.60\textwidth}{!}{
  \begin{tabular}{l|cc|cc}
    \toprule
     Method &  CER ($\downarrow$)  & WER ($\downarrow$) & EER ($\downarrow$) & SECS ($\uparrow$) \\
    \midrule
       GT & 0.28& 0.74 & - & - \\
       \midrule
    DDDM-TTS  (Ours) &  0.71& 0.95 & 5.00 &0.824\\
    \bottomrule
  \end{tabular}
  }

\end{table*}

\end{document}